\documentclass[aps,prd,superscriptaddress,onecolumn,floatfix,nofootinbib,preprintnumbers]{revtex4}

\usepackage{amsmath,graphicx,tabularx,url}

\begin{document}

\def\tablename{Table}
\def\figurename{Figure}

\newcommand{\heptop}{{\sc HEPTopTagger \,}}

\newcommand\one{\leavevmode\hbox{\small1\normalsize\kern-.33em1}}
\newcommand{\Mpl}{M_\mathrm{Pl}}
\newcommand{\p}{\partial}
\newcommand{\lag}{\mathcal{L}}
\newcommand{\qqquad}{\qquad \qquad}
\newcommand{\qqqquad}{\qquad \qquad \qquad}

\newcommand{\qb}{\bar{q}}
\newcommand{\matx}{|\mathcal{M}|^2}
\newcommand{\really}{\stackrel{!}{=}}
\newcommand{\msbar}{\overline{\text{MS}}}
\newcommand{\qns}{f_q^\text{NS}}
\newcommand{\lqcd}{\Lambda_\text{QCD}}
\newcommand{\met}{\slashchar{p}_T}
\newcommand{\pmiss}{\slashchar{\vec{p}}_T}

\newcommand{\st}[1]{\tilde{t}_{#1}}
\newcommand{\stb}[1]{\tilde{t}_{#1}^*}
\newcommand{\nz}[1]{\tilde{\chi}_{#1}^0}
\newcommand{\cp}[1]{\tilde{\chi}_{#1}^+}
\newcommand{\cm}[1]{\tilde{\chi}_{#1}^-}

\providecommand{\mg}{m_{\tilde{g}}}
\providecommand{\mst}{m_{\tilde{t}}}
\newcommand{\msn}[1]{m_{\tilde{\nu}_{#1}}}
\newcommand{\mch}[1]{m_{\tilde{\chi}^+_{#1}}}
\newcommand{\mne}[1]{m_{\tilde{\chi}^0_{#1}}}
\newcommand{\msb}[1]{m_{\tilde{b}_{#1}}}

\newcommand{\mev}{{\ensuremath\rm MeV}}
\newcommand{\gev}{{\ensuremath\rm GeV}}
\newcommand{\tev}{{\ensuremath\rm TeV}}
\newcommand{\fb}{{\ensuremath\rm fb}}
\newcommand{\ab}{{\ensuremath\rm ab}}
\newcommand{\pb}{{\ensuremath\rm pb}}
\newcommand{\sign}{{\ensuremath\rm sign}}
\newcommand{\ifb}{{\ensuremath\rm fb^{-1}}}

\def\slashchar#1{\setbox0=\hbox{$#1$}           
   \dimen0=\wd0                                 
   \setbox1=\hbox{/} \dimen1=\wd1               
   \ifdim\dimen0>\dimen1                        
      \rlap{\hbox to \dimen0{\hfil/\hfil}}      
      #1                                        
   \else                                        
      \rlap{\hbox to \dimen1{\hfil$#1$\hfil}}   
      /                                         
   \fi}
\newcommand{\dslash}{\slashchar{\partial}}
\newcommand{\Dslash}{\slashchar{D}}

\def\eg{{\sl e.g.} \,}
\def\ie{{\sl i.e.} \,}
\def\etal{{\sl et al} \,}

\title{How to Improve Top Tagging}

\author{Tilman Plehn}
\affiliation{Institut f\"ur Theoretische Physik, 
             Universit\"at Heidelberg, Germany}

\author{Michael Spannowsky}
\affiliation{Institute for Particle Physics Phenomenology, Department of Physics, Durham University, DH1 3LE, United Kingdom}

\author{Michihisa Takeuchi}
\affiliation{Institut f\"ur Theoretische Physik, 
             Universit\"at Heidelberg, Germany}

\begin{abstract}
  In time for the first tests on LHC data we introduce a set of
  improvements and tests of purely kinematic top tagging
  algorithms. First, we show how different jet algorithms can be used
  for different transverse momentum regimes. Combining pruning and
  filtering in the reconstruction can enhance the signal over
  background ratio significantly, while larger jet radii only give
  minor improvements. Finally, bottom tagging can be added to the top
  tagger, but at least for the {\sc HEPTopTagger} does not improve the
  kinematic selection algorithm.
\end{abstract}

\maketitle

\section{Introduction}
\label{sec:intro}

The top quark, found in 1995~\cite{topdiscovery}, is the heaviest and
so-far only observed fermion with a weak-scale mass. Therefore, it is
expected to have strong ties to the mechanism which triggers
electroweak symmetry breaking. Searches for new physics in the top
sector are of high priority because they can shed additional light on
the structure of the Standard Model at and above the weak scale.  Many
extensions of the Standard Model, like supersymmetry or little Higgs
models~\cite{review}, predict top partners to ameliorate the
loop-induced effect of the top quark on the Higgs-boson's mass.
Typical signatures for such extended top sectors include top partners
decaying to a top quark and missing
energy~\cite{meade,heptop,semilep,stops_measure} or heavy resonances
decaying to two often strongly boosted top
quarks~\cite{semi_resonances,had_resonances}.

During the Tevatron's final years D0 and CDF have measured several
anomalies directly related to the top sector: in CDF's single top
analysis the ratio between the $s$- and $t$-channel production rates
deviates by $2.5 \sigma$ from the SM
prediction~\cite{Aaltonen:2010jr}, and both experiments measure an
enhanced $t\bar{t}$ forward-backward asymmetry compared to SM
predictions~\cite{Aaltonen:2011kc}. Also including the excess in the di-jet
invariant mass spectrum of the $Wjj$ final state, measured (only) by
CDF~\cite{Aaltonen:2011mk}, all these anomalies show that we need an
improved understanding and simulation of top
production processes~\cite{wjets}.

The top pair production rate at the LHC ranges around one million tops
per inverse femtobarn of integrated luminosity.  On the one hand, this
means that top pair production is a very challenging background for
searches relying on high multiplicity final states of jets, leptons
and missing energy~\cite{meade,semilep,heptop,stuff}. On the other
hand, this means that already now we can test top pair events in many
different kinematic regimes. In this paper we will focus on moderately
boosted top quarks in the semileptonic decay channel.\bigskip

While the idea of studying the substructure of jets is already a
classic~\cite{seymour}, the potential for searches of massive Higgs
and gauge bosons has only been appreciated
recently~\cite{bdrs,substructure,subreviews}.  In this paper we will focus on
tagging boosted top
quarks~\cite{early,hopkins,template,tth,pruning,scetfact,trimming,heptop,scet,recent}. Aside
from being sensitive probes of new physics they are also the prime
candidates to generally establish that fat-jet or subjet methods work
at the LHC.  Some very promising ATLAS results on the {\sc
  HEPTopTagger} ~\cite{tth,heptop} performance on data can be found in
Ref.~\cite{talkgregor}. CMS has already released first search results
using a top tagger~\cite{cms}.  Of the Tevatron anomalies listed above
the top forward-backward or charge asymmetry is particularly
interesting in the light of boosted top quarks; the ratio of initial
state quarks vs gluons increases in the boosted regime, thereby
enhancing the otherwise small asymmetry at the LHC~\cite{Afb}.\bigskip

Starting from the default purely kinematic setup of the {\sc
  HEPTopTagger} we investigate several avenues on how to improve its
performance: In Sec.~\ref{sec:reco} we discuss how well the momentum
of subjets matches the decay partons in the default setup and which
strategies for an improvement should be promising.  In
Sec.~\ref{sec:algos} we investigate the performance of different jet
algorithms for the filtering and subjet reconstruction.  In
Sec.~\ref{sec:prune} we then study the tagging performance if we
include pruning in combination with filtering.  The pruned top mass we
use as an additional kinematic variable according
to Ref.~\cite{soper_spanno}. In Sec.~\ref{sec:size} we investigate the
possibility to enlarge the size of the fat jet to $R=1.8$, focussing
on the currently most relevant low-$p_T$ tops.  Finally, in
Sec.~\ref{sec:btag} we augment the kinematic top tagger by a $b$-tag
inside the fat jet~\cite{giacinto}. Two possible strategies are simply
adding the $b$-tag at the end of the top tagging algorithm or
including it in a modified extraction of the relevant subjets.

\section{Subjet-parton reconstruction}
\label{sec:reco}

Before we can suggest and test improvements to our top tagger it is
crucial that we study measures for the quality of the top
reconstruction.  The geometrical distance between the reconstructed
and the true top momenta is simply
\begin{equation}
\Delta R^2_\text{top} 
= \Delta R^2(p_\text{top}^\text{tagged},p_\text{top}^\text{parton}) \; .
\label{eq:rtop}
\end{equation}
For a more detailed study we also compute the geometric separation of
the top decay products, which requires a proper definition of the
parton and jet level constituents. Jet combinatorics is the main
challenge, in particular for hadronic top pair production at the
LHC. For example including up to two additional hard QCD jets the
$t\bar{t}$ sample consists of 6 to 8 partons which we label as (1,4)
for the bottom quarks, (2,5) for the harder $W$ decay partons, (3,6)
for the softer $W$ decay partons, and (7,8,...) for additionally
radiated partons. The partons 1-3 and 4-6 come from one top decay
each.  After hadronization and jet reconstruction the corresponding
$b$, $W_1$, and $W_2$ subjets are defined such that ${W_1}$ (harder)
and ${W_2}$ (softer) reconstruct $m_W$ best. Note that we do not apply
any $b$-tagging, an issue we will look at in Sec.~\ref{sec:btag}. We
can then define
\begin{equation}
\Delta R_\text{sum}^2
= \min_\text{mappings} \sum_{i=1}^3 \Delta R^2(p_i^\text{sub},p_{j_i}^\text{parton}) \; ,
\label{eq:rsum}
\end{equation}
where the label $j_i$ denotes the $i$-th hardest parton in the tagged
top. The best parton-subjet mapping $\{j_i\}=\{j_1,j_2, j_3\}$ is
defined by the minimum $\sum_{i=1}^3 \Delta
R^2(p_i^\text{sub},p_{j_i}^\text{parton})$ value.  For example, for
$\{j_i\}=\{1,5,7\}$ the hardest subjet corresponds to a $b$-quark, the
second hardest to a $W$ decay from the other top, and the softest an
additional parton from jet radiations. This way we can categorize all
tagged tops into three types:
\begin{itemize}
\item[] type~1: $\{ j_1, j_2, j_3 \}$ come from one top decay, \ie
  $\{1,2,3\}$ or $\{4,5,6\}$.
\item[] type~2: only the two hardest $\{ j_1, j_2\}$ come from one top
  decay, $j_3$ has a different origin.
\item[] type~3: else.
\end{itemize}
\bigskip
%

\begin{figure}[b]
\includegraphics[width=0.32\textwidth]{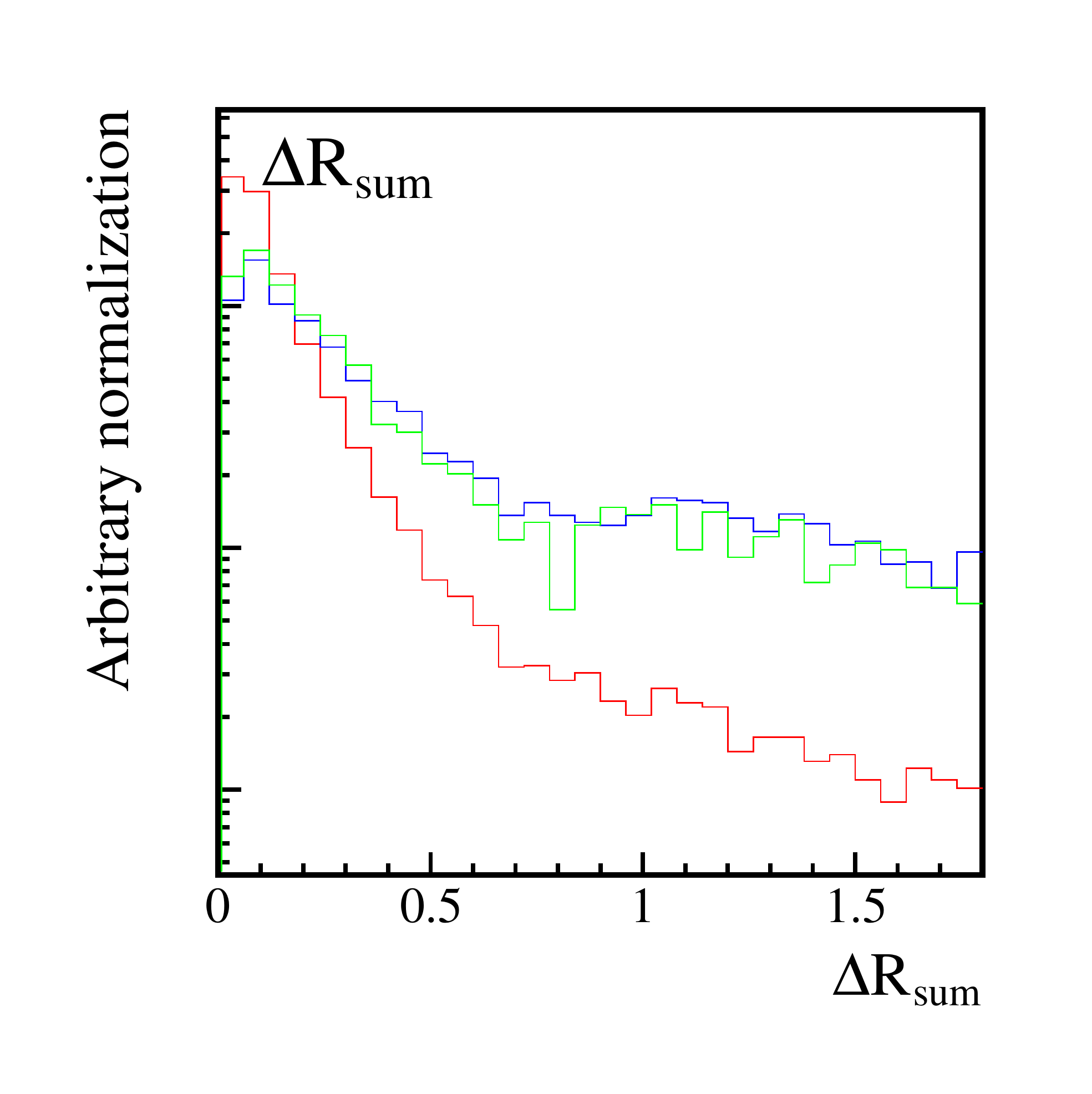}
\includegraphics[width=0.32\textwidth]{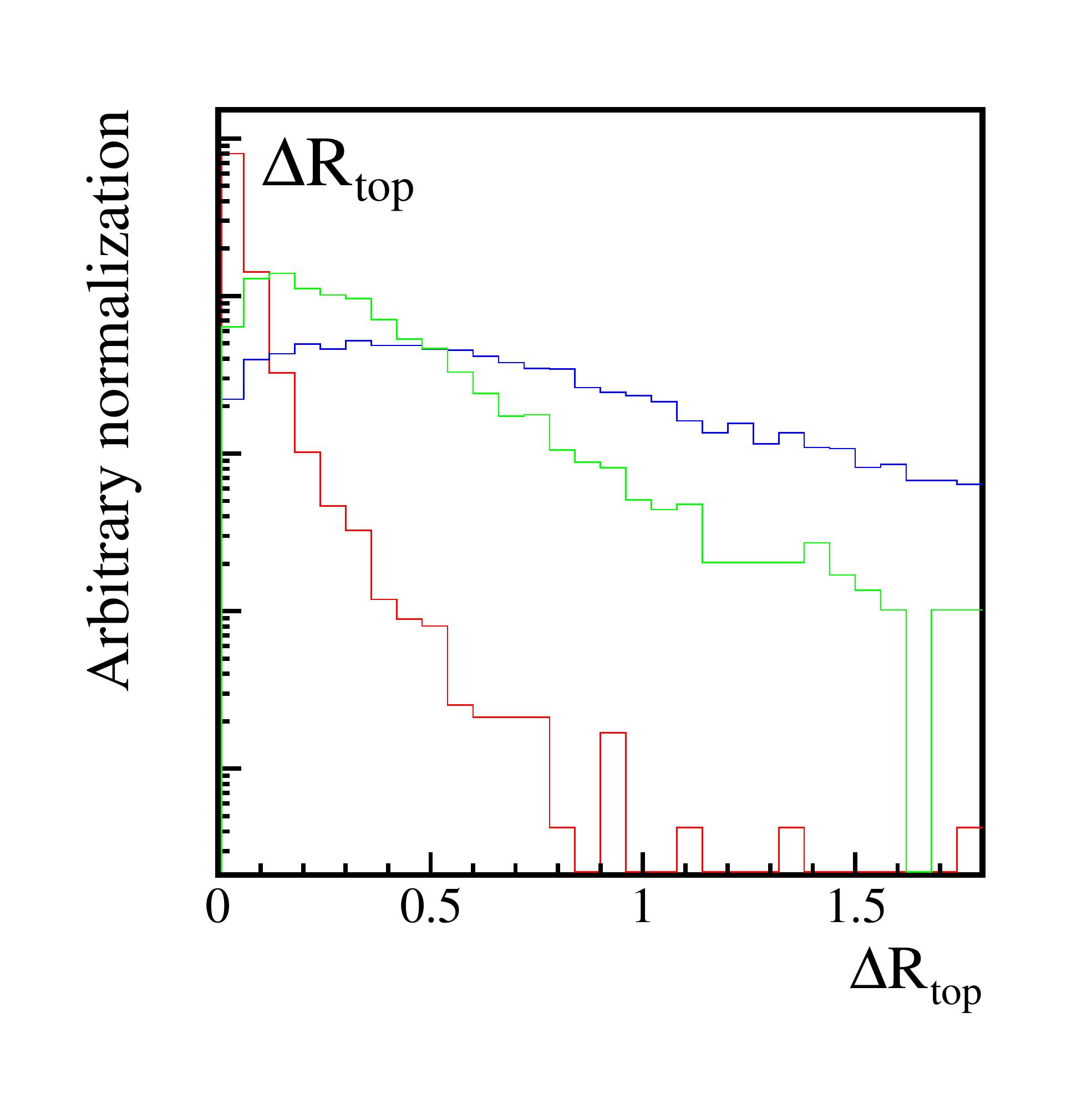}
\includegraphics[width=0.32\textwidth]{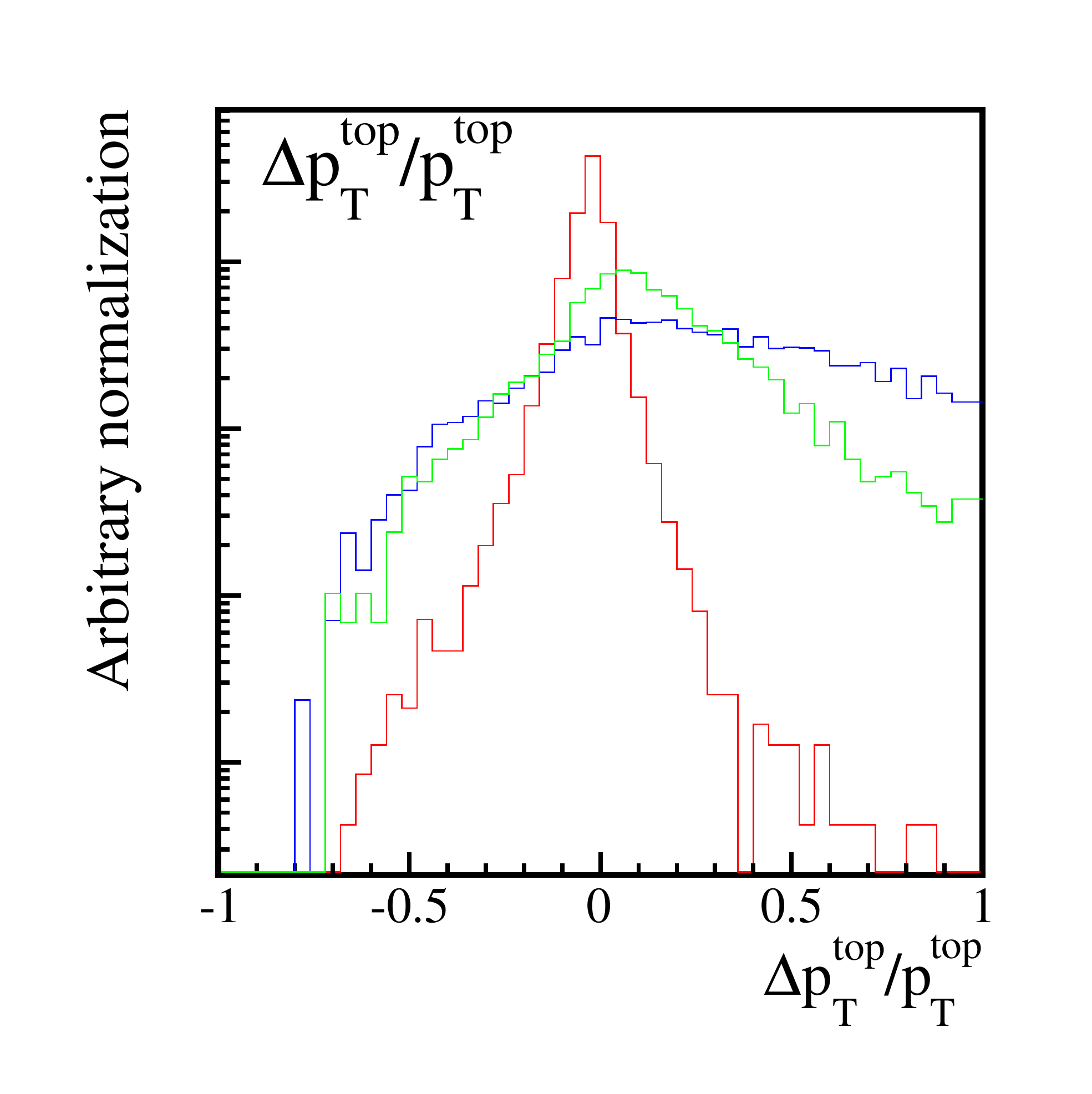}
\vspace*{-4mm}
\caption{Left: $\Delta R_\text{sum}$ as defined in Eq.\eqref{eq:rsum};
  center: $\Delta R_\text{top}$ as defined in Eq.\eqref{eq:rtop};
  right: $\Delta p_T^\text{top}/p_T$. The colors represent type~1
  (red), type~2 (green), type~3 (blue) for semileptonic top pairs at a
  collider energy of 14~TeV.}
  \label{fig:1}
\end{figure}

For semileptonic top pairs the combinatorics is simplified significantly,
but we can still categorize all tagged tops along the same lines.  Our
analysis is based on {\sc Alpgen-Pythia}~\cite{alpgen,pythia} samples
with MLM merging~\cite{mlm} ($R^\text{MLM}=0.4, p_T^\text{MLM}>30$~GeV) and we cluster the visible final state using {\sc FastJet}~\cite{fastjet}. We take
into account two hard jets in association with $t\bar{t}$ production
and three to five hard jets for $W$+jets and QCD jets. The $t\bar{t}$
sample we re-weight to 918~pb~\cite{top_rate}. The left panel of
Fig.~\ref{fig:1} shows $\Delta R_\text{sum}$ for each
type. The quality of the reconstructed subjets direction is the same
for all types, with the exception of long type~2 and type~3 tails.  In
the central and right panels of Fig.~\ref{fig:1} we test the actual top
momentum reconstruction in terms of $\Delta R_\text{top}$ and $\Delta
p_T^\text{top}/p^\text{top}_T = (p_{T,t}^\text{tagged} -
p_{T,t}^\text{parton})/p_{T,t}^\text{tagged}$.  Its quality depends on
the different types of parton identification and the most poorly
reconstructed candidate tops are of type~2 and type~3. Unlike for
type~1 tops their distributions do not follow a largely Gaussian shape
centered at zero but show a significant shift.

\begin{figure}[t]
\includegraphics[width=0.32\textwidth]{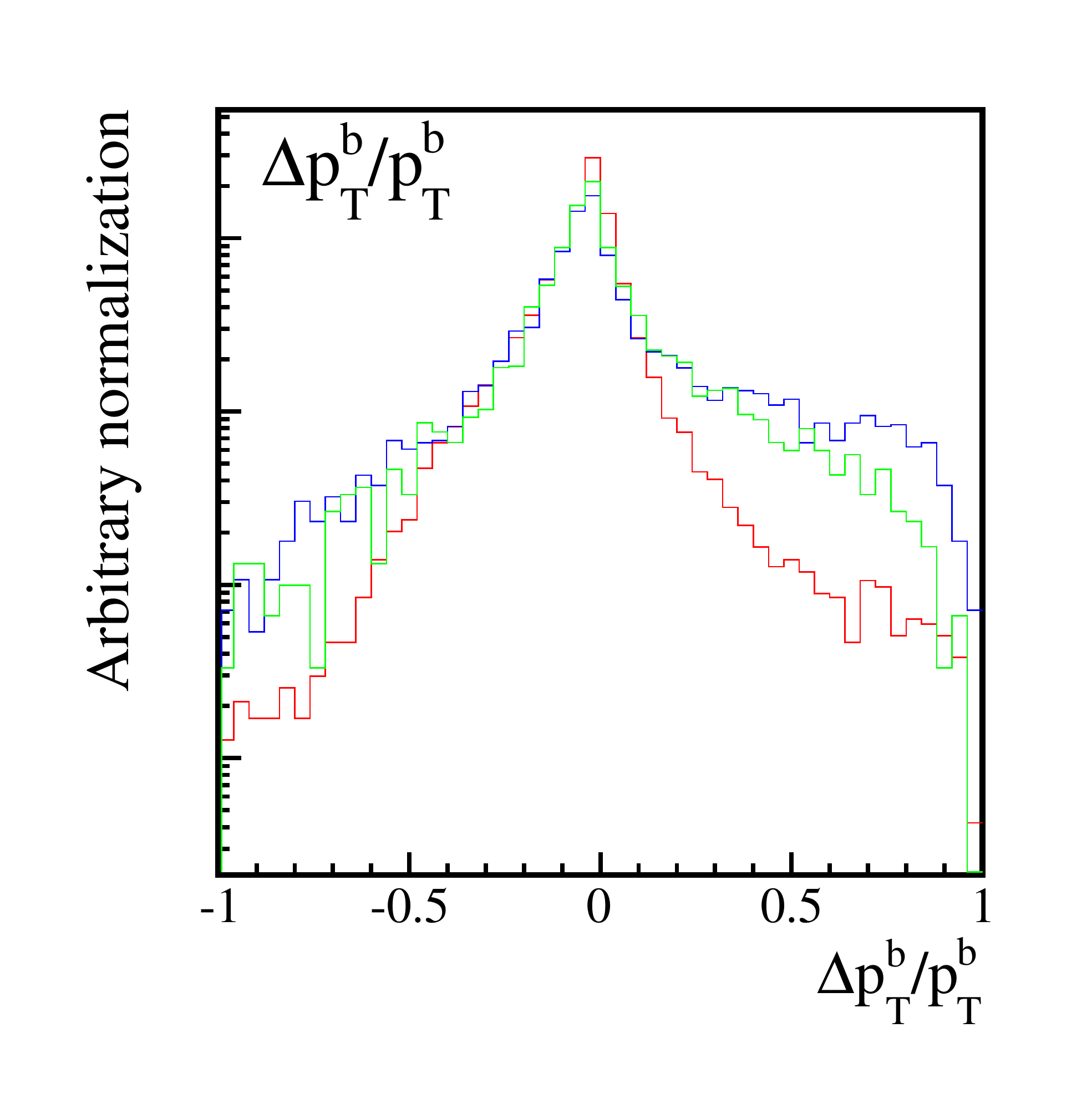}
\includegraphics[width=0.32\textwidth]{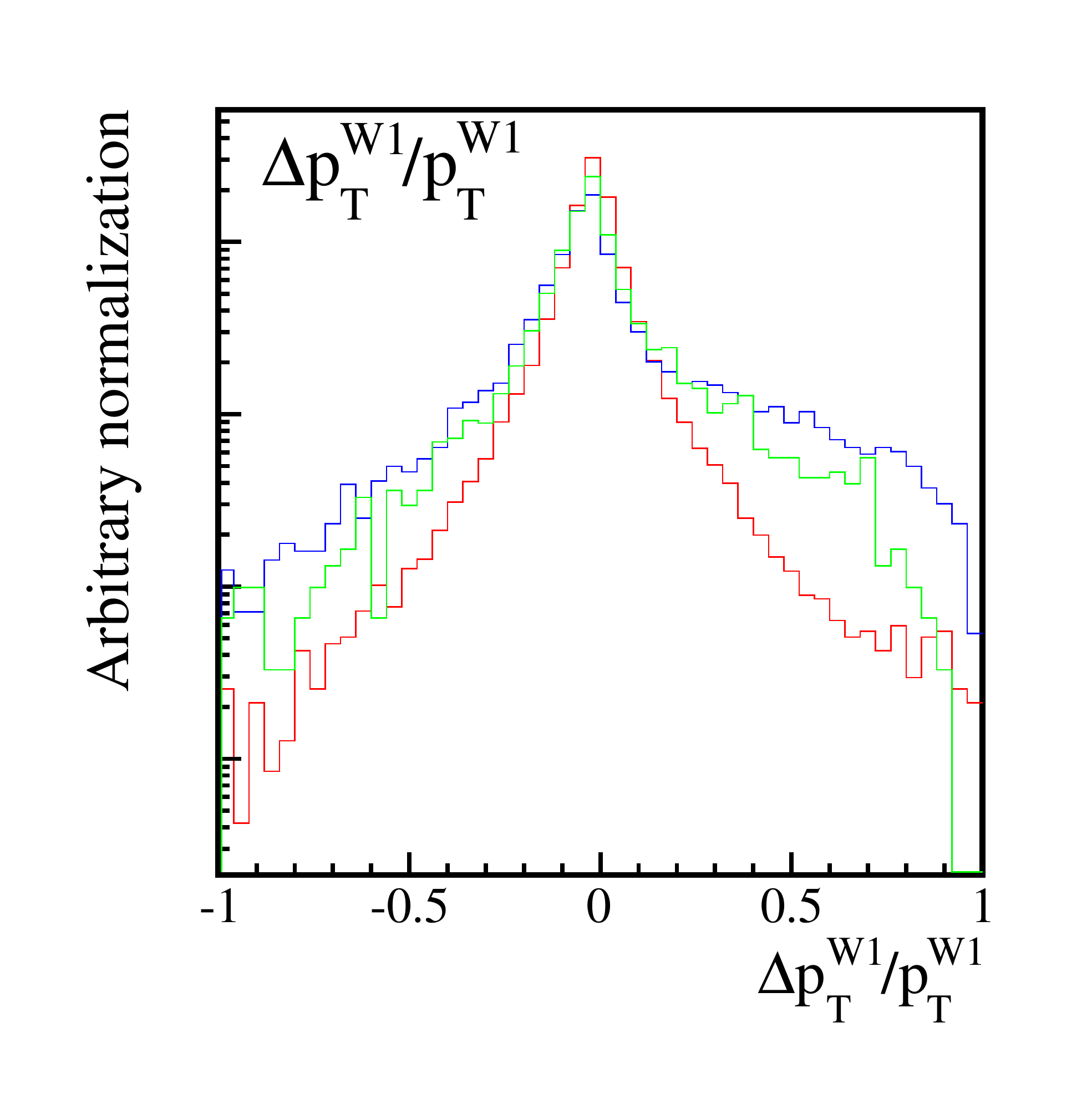}
\includegraphics[width=0.32\textwidth]{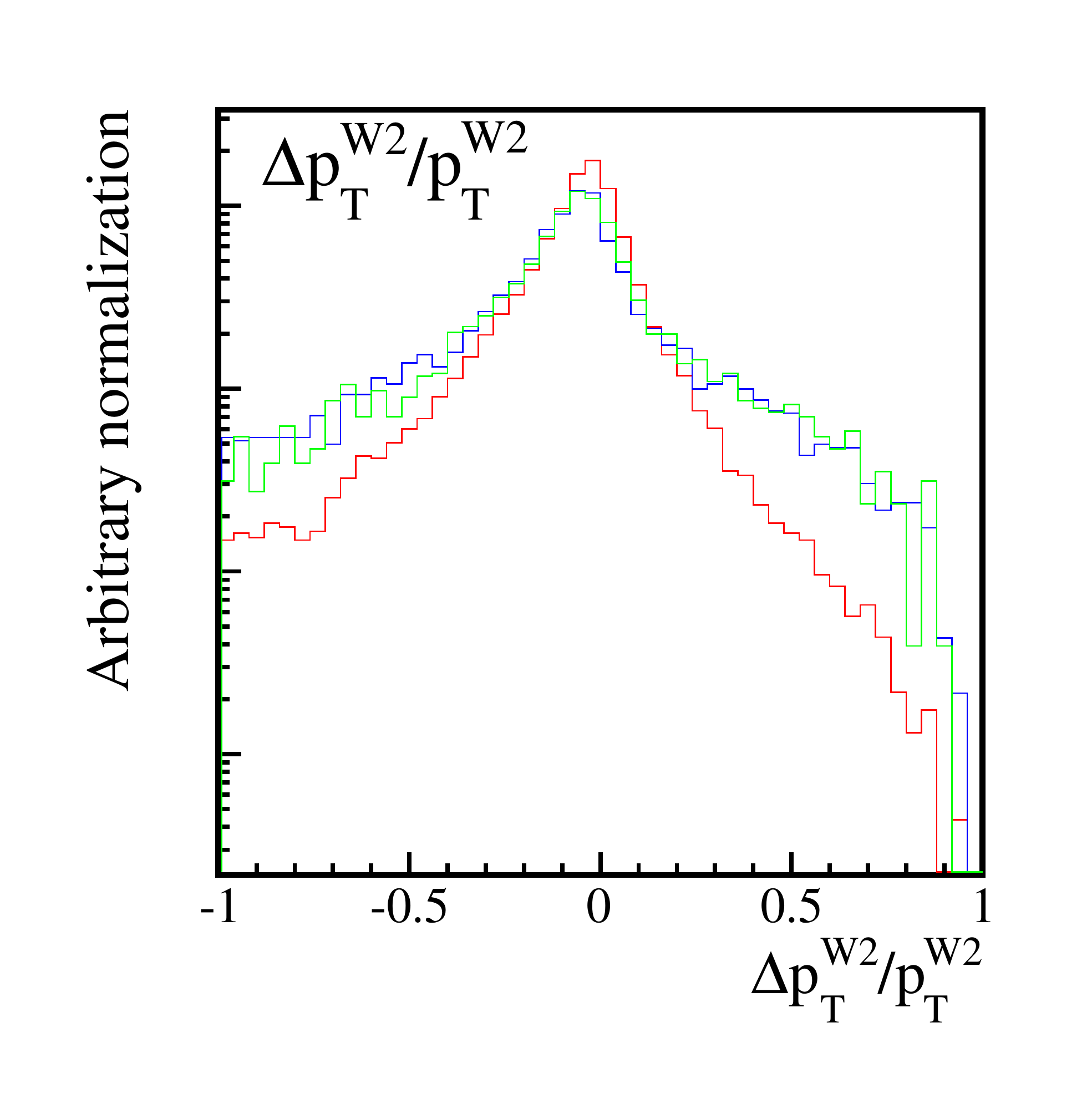}
\vspace*{-4mm}
\caption{Relative $p_T$ differences between subjets and the
  corresponding partons $\Delta p_T/p_T$ for the $b$-subjet (left),
  the $W_1$-subjet (center), and the $W_2$ subjet (right). Again, the
  different colors show type~1 (red), type~2 (green) and type~3
  (blue).}
 \label{fig:2}
\end{figure}

Fig.~\ref{fig:2} shows the transverse momentum difference between each
subjet and the corresponding parton $\Delta p_T/p_T =
(p_T^\text{subjet} - p_T^\text{parton})/p_T^\text{subjet}$.  We see
that the subjet momentum reconstruction is essentially of the same
quality for all subjets and for all types, with the exception of
$p_T^{W1}$ which is better reconstructed than $p_T^{W2}$ because of
its larger value.

\begin{table}[b]
\begin{small}
\begin{tabular}{c|rrrr||rrrr}
\hline
&\multicolumn{4}{c||}{all $p_T^\text{tagged}$}
&\multicolumn{4}{c}{$p_T^\text{tagged}>250$~GeV}\cr
\hline
&tagged &$\Delta R_\text{sum} < 0.4$ & $\ \Delta R_\text{top} < 0.2$ &$\ \left|\dfrac{\Delta p_T^{\text{top}}}{p_T} \right|< 0.15$
&tagged &$\Delta R_\text{sum} < 0.4$ & $\ \Delta R_\text{top} < 0.2$ &$\ \left|\dfrac{\Delta p_T^{\text{top}}}{p_T} \right|< 0.15$\cr
\hline
total & 14156 &11904 (84\%)&10841 (77\%)&11037 (78\%)&6029&5279 (88\%)&5191 (86\%)&5170 (86\%)\cr
type~1 &10318 &9531 (92\%)&10102 (98\%)&9897 (96\%)&4919 &4624 (94\%)&4858 (99\%)&4774 (97\%)\cr
type~2 & 1336 &896 (67\%)&477 (36\%)&623 (47\%)&412 &273 (66\%)&218 (53\%)&244 (59\%)\cr
type~3 & 2503 &1478 (59\%)&263 (11\%)&517 (21\%)&698&381 (55\%)&115 (16\%)&152 (22\%)\cr
\hline
\end{tabular}
\end{small}
\caption{Tagged top rates (in fb) in the semileptonic $t\bar{t}$
  sample for 14~TeV collider energy.  The percentages for the
  additional cuts are relative to the numbers of tagged tops in each
  row.}
\label{tab:1}
\end{table}

In Tab.~\ref{tab:1} we give the fraction of tagged tops for each of
the three types after different requirements on the quality of the
reconstruction.  For type~1 tags the momentum reconstruction both for
subjets and tops is almost perfect.  For type~2 and type~3 tags, a large
fraction of tagged tops satisfies $\Delta R_\text{sum}<0.4$, which
means the individual subjets are reconstructed well but the set of the
partons is wrongly picked. Consequently, the top momentum
reconstruction for type~3 tags becomes worse. Thanks to a correct
assignment for the hardest two subjets in type~2 tags the top momentum
reconstruction is not too bad because the wrong third subjet does not
contribute much to the top momentum.\bigskip

From the discussion above we can conclude that the individual
subjet-parton momentum reconstruction works well for all types. The
limitation to the top momentum reconstruction arises from events where
some of the identified subjets do not correspond to a top decay
product.  From Tab.~\ref{tab:1} we estimate that $\mathcal{O}(20\%)$ of
type~3 tops and $\mathcal{O}(50\%)$ of type~2 tops still give the correct
top momentum within 15\%. The fraction of tagged tops with good
momentum reconstruction within each type does not depend much on
$p_{T,t}$; however, the fraction of type~2 and type~3 tags in all
tagged tops decreases for higher $p_T$ and effectively leads to a
better momentum reconstruction.  In total, 85\% of all tagged tops and
up to 90\% of all tags with $p_T^\text{tagged}>250$~GeV reproduce the
true momentum within a 15\% error bar.\bigskip

An assigned top tagging efficiency should describe which fraction of
hadronic tops in any event sample are tagged.  Such an efficiency we
define step by step:
\begin{enumerate}
\item all decay products satisfy \underline{$R_\text{C/A}
  <1.5$}. 
\item all decay products appear in a \underline{fat jet}, \ie there
  exist unfiltered subjets with $\Delta R(p^\text{parton}, p^\text{subjet})
  <0.4$.
\item all decay products appear in a fat jet with a top
  \underline{candidate} fulfilling $150 < m_{jjj}^\text{filter} <200$~GeV.
\item all decay products appear in a fat jet with a \underline{tagged}
  top, \ie after the mass plane cut.
\end{enumerate}
For the first step we use $R_\text{C/A}=\max\{R_1,R_2\}$ based on the
two C/A measures for the necessary clustering steps $R_{1,2}$.  All
these events are defined for the signal, so we can show them as a
function of the true $p_{T,t}$ at parton level.\bigskip

\begin{figure}[b]
\includegraphics[width=0.32\textwidth]{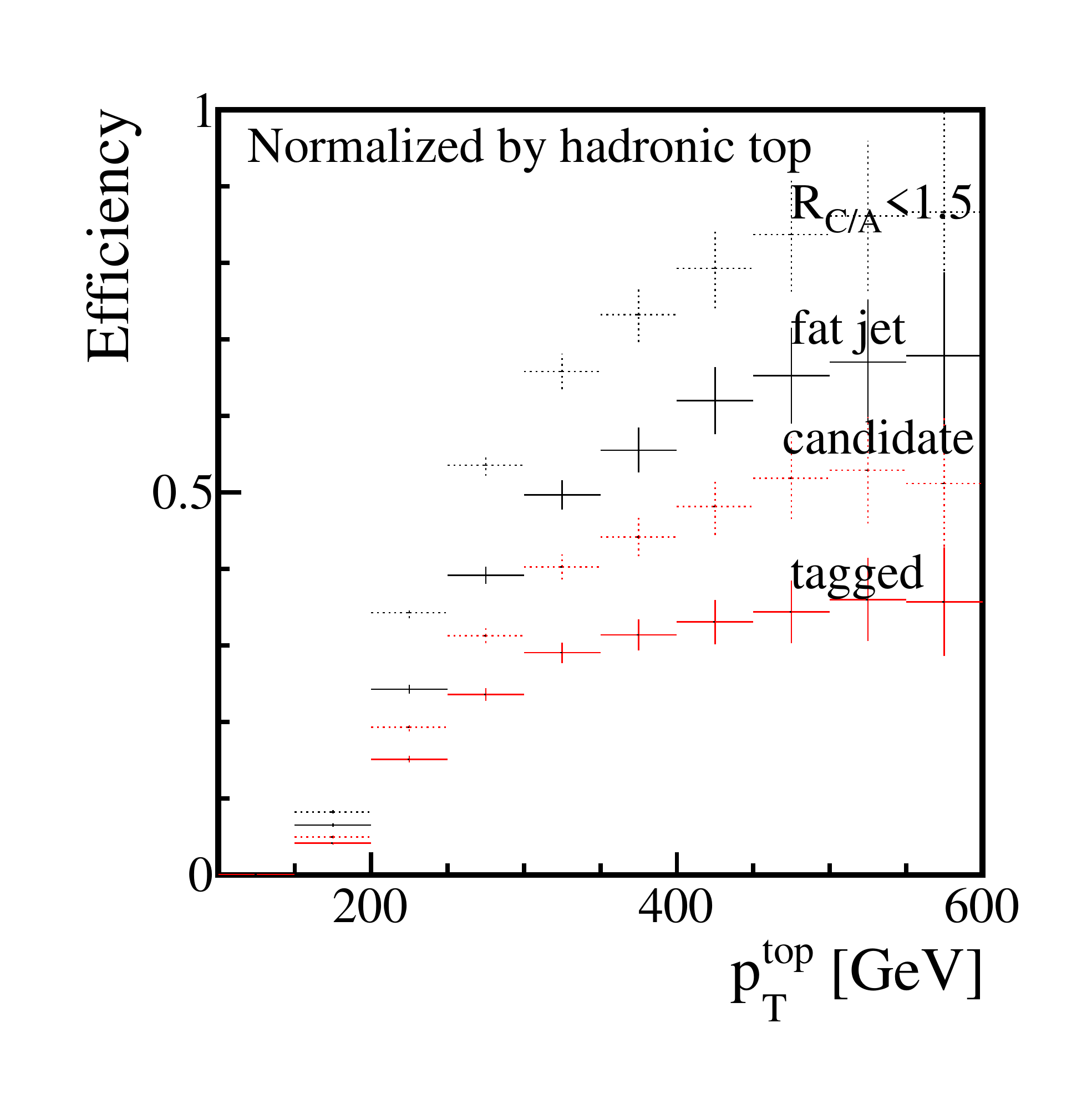}
\includegraphics[width=0.32\textwidth]{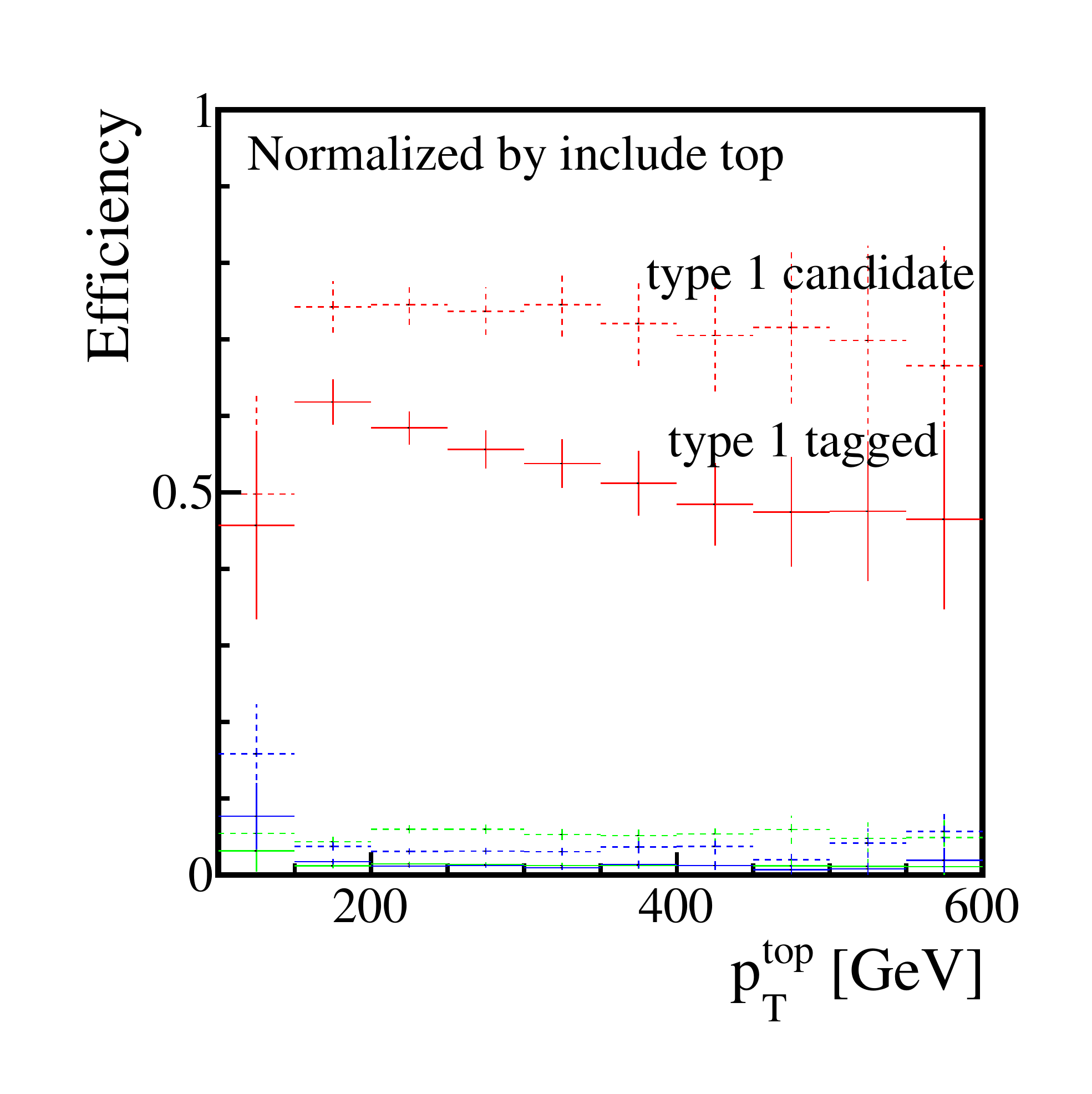}
\vspace*{-4mm}
\caption{Left: tagging efficiencies normalized to the number of
  hadronic tops as a function of $p_T$.  Right: tagging efficiencies
  relative to the number of hadronic tops included in a fat jet. The
  different curves are discussed in the text.}
\label{fig:had}
\end{figure}

The left panel of Fig.~\ref{fig:had} shows the event fractions
corresponding to all four categories normalized by the number of
hadronic tops as a functions of $p_T$. First, the dotted entries show
the fraction of tops with $R_\text{C/A}<1.5$, while the solid entries
show the fraction of top decay products inside a fat jet.  For any fat
jet analysis they fix an upper bound on all tagging efficiencies. 

The red dotted and solid entries show the fraction of candidate and
tagged top events, all constrained to type~1 tags.  The difference
between the two is simply given by the mass plane cuts.  After these
cuts roughly 30\% of all hadronic tops are tagged above $p_T \sim
250$~GeV.  Note that there exist type~2 and type~3 candidates and
tags, \ie hadronic tops whose decay products are included in a fat jet
but the extracted subjets do not fall into type~1.  

The right panel of Fig.~\ref{fig:had} first shows the fraction of
type~1 candidates and type~1 tags relative to the number of fat jets
including all three top decay products, corresponding to the second
category in the above list.  Type~2 and type~3 are shown in green and
blue.  The difference between all top decay products inside a fat jet and
type~1 candidates is about 25\% and almost constant for $200 < p_T <
500$ GeV. It is partly (at most 10\%) due to the existence of type~2
or three candidates, which means the tagger wrongly selects subjets even
though the fat jet does include all decay products. Alternatively,
there might be overlapping of subjets such that any three subjets are
inconsistent with the top mass constraint.  The difference between
candidates and tags corresponds to the mass plane cuts, where some
signal loss is inevitable for rejecting QCD and $W+$jets
backgrounds. The numbers of type~2 and type~3 tags are negligible;
most of the type~2 and type~3 tags shown in Tab.~\ref{tab:1} come from
fat jets not including a top. The mass plane cuts efficiently remove
such contributions.  Hence, we see that once the top decay products
are captured in a fat jet, 50\% to 60\% of the tops are tagged.

The rapid drop of efficiency below $p_T \sim 250$~GeV in the left
panel of Fig.~\ref{fig:had} is simply due to the rapid drop of the
fraction of decay products within $R_\text{C/A}<1.5$ and inside the
fat jet. The ratio of type~1 tags to decay products within a fat jet
is not small for $150 < p_T < 250$~GeV. Therefore, we will test
an increased $R$ value in Sec.~\ref{sec:size}.

An obvious question from the right panel of Fig.~\ref{fig:had} is why
the tagging efficiency degrades for larger $p_T$, where in principle
the tagging performance should tend to increase.  This effect of the
mass plane cut is caused by a mis-reclustering after filtering and is
more pronounced for the C/A algorithm. The C/A algorithm relies on the
$R$ distance exclusively, therefore the softest two of the five filtered subjets are not
necessarily combined according to their shower history.  As a result, we find unbalanced
invariant masses from the three re-clustered subjets.  This happens more
frequently in the high-$p_T$ regime where all five subjets are not well
separated, so the rejection probability by the mass plane cut
increases with $p_T$.  This tendency and the possibility of changing
the underlying jet algorithm and its effect on mass plane cuts we will
discuss in Sec.~\ref{sec:algos}.\bigskip

To include backgrounds and mis-tagging efficiencies we need to define
a slightly different set of scenarios, namely as a function of the fat
jet $p_T$. We define four scenarios similar to the ones before, but now
in terms of fat jets:
\begin{enumerate}
\item fat jets with \underline{three subjets} or more after the
  mass drop criteria
\item \underline{fat jets} where all distances between top decay
  products and their closest subjets are less than $0.4$.
\item fat jets with a top \underline{candidate}
\item fat jets with a top \underline{tag} 
\end{enumerate}

\begin{figure}[b]
\includegraphics[width=0.32\textwidth]{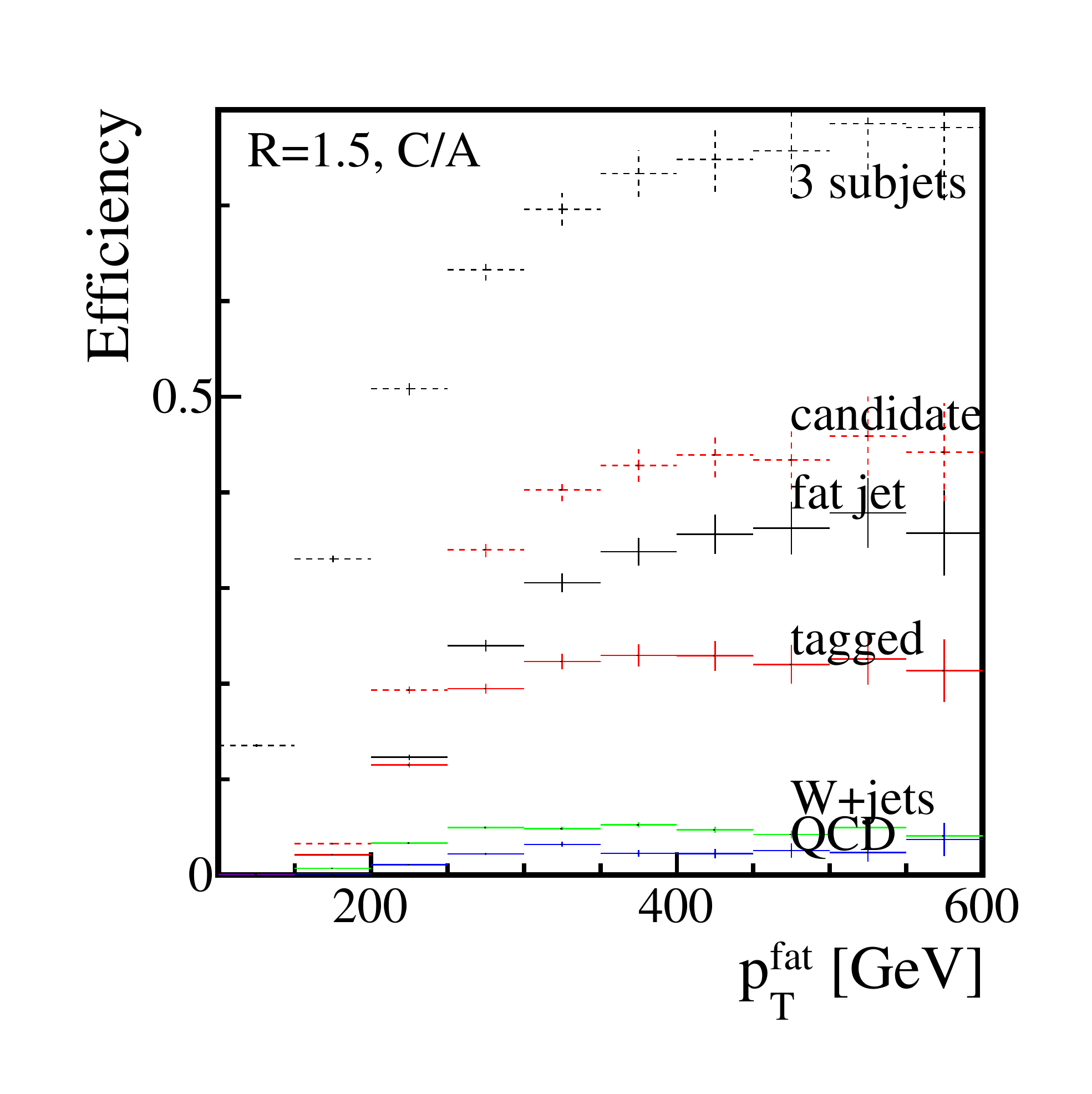}
\includegraphics[width=0.32\textwidth]{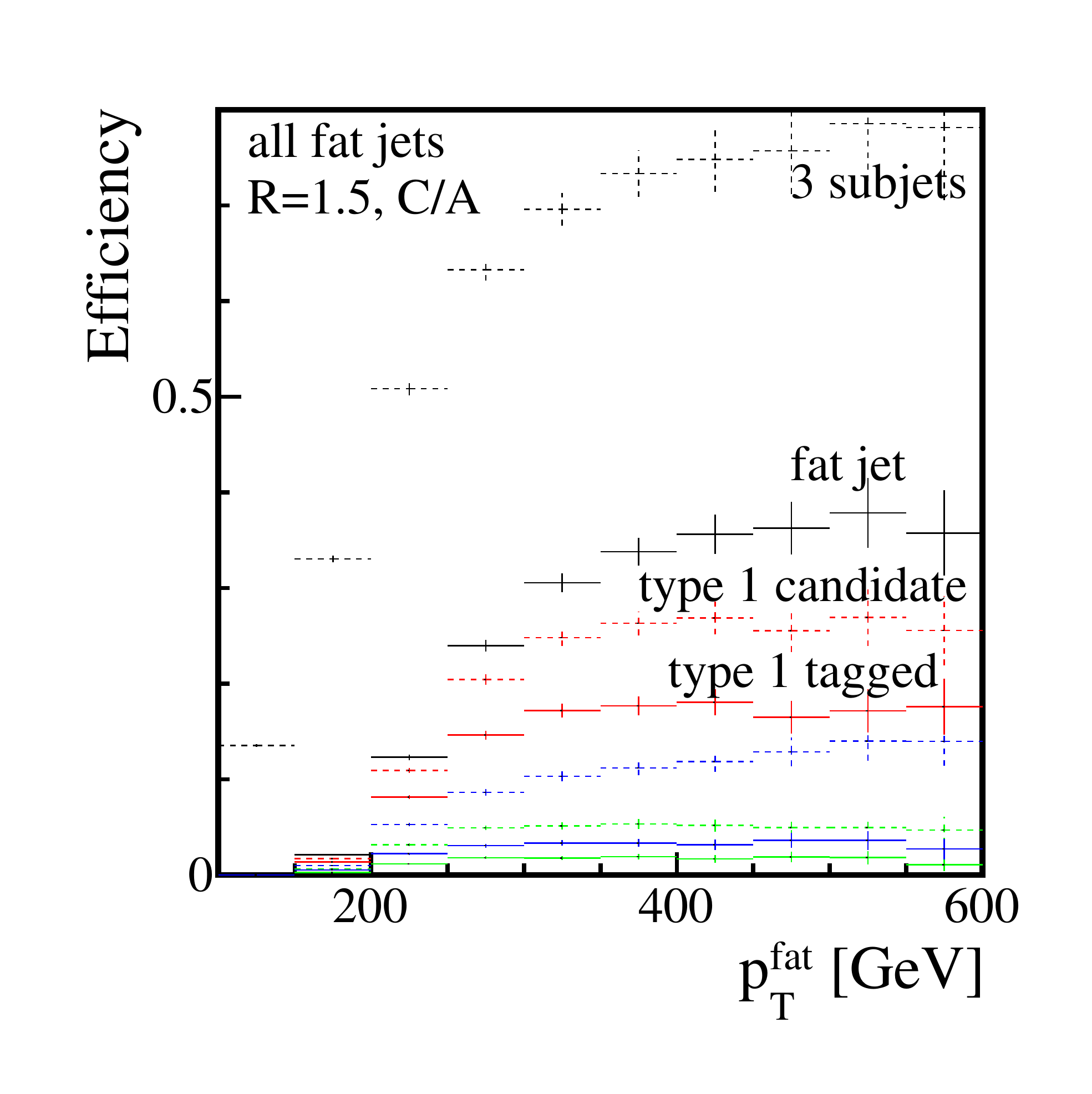}
\includegraphics[width=0.32\textwidth]{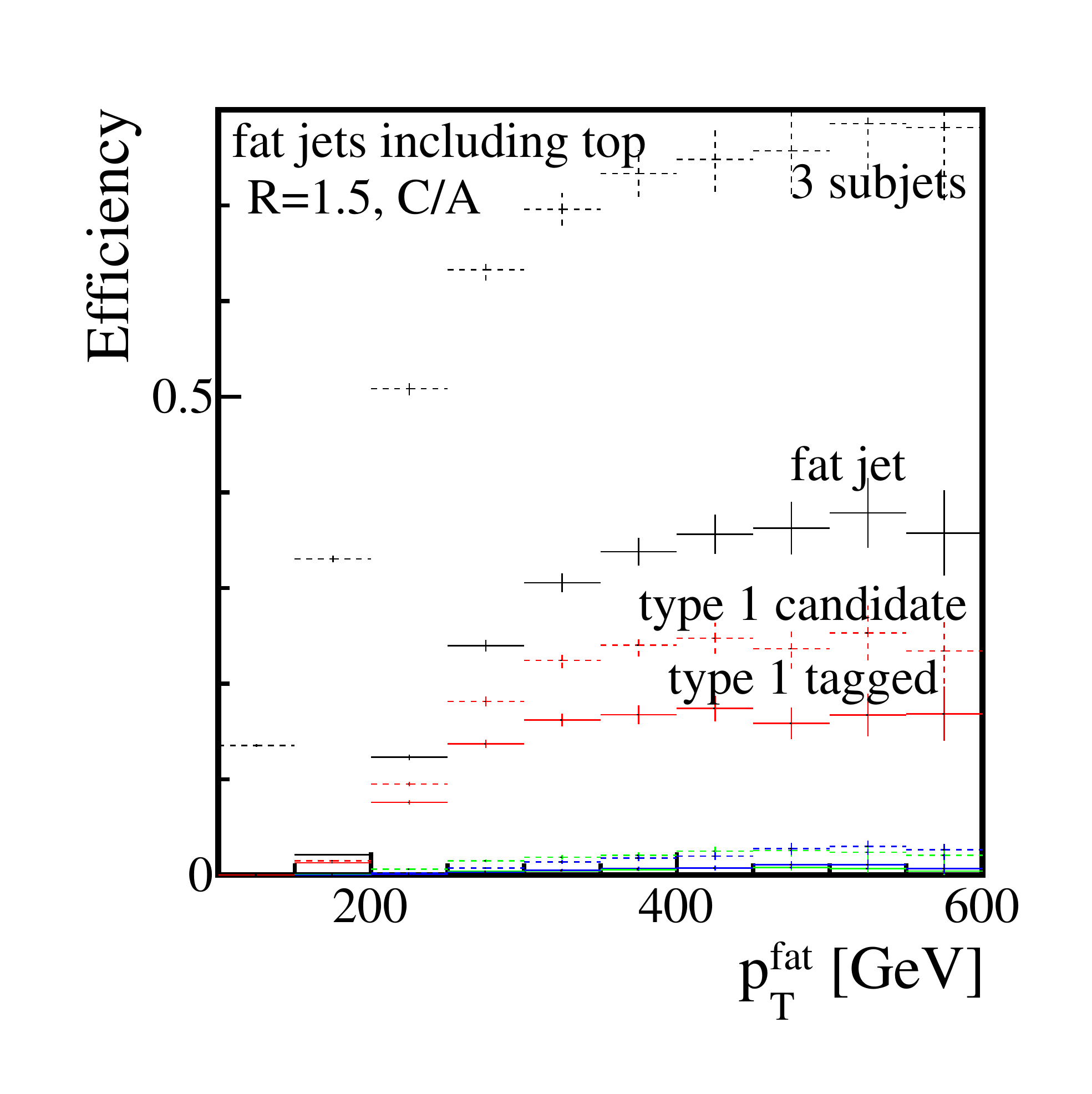}
\vspace*{-4mm}
\caption{ Left: efficiencies $\epsilon_{t\bar{t}}$,
  $\epsilon_{W+\text{jets}}$ $\epsilon_\text{QCD}$ as functions of
  the fat jet $p_T$.  Center: fraction of tagged tops for type~1
  (red), type~2 (green) and type~3 (blue).  The dotted lines show the
  corresponding candidate fractions.  Right: fraction of type~1,
  type~2 and type~3 only in for fat jets including a top.}
  \label{fig:fat}
\end{figure}

The left panel of Fig.~\ref{fig:fat} shows all corresponding fractions
relative to the number of fat jets as a function of $p_T$. The black
dotted and solid symbols show the fraction of events with at least
three subjets and of all top decay products included.  More than
half of the fat jet events with at least three subjets do not
include the top decay products, even for the semileptonic $t\bar{t}$
sample. The red entries show the fraction of candidates (dotted) and
tags (solid).  Because all efficiencies are shown as a function of the
fat jet $p_T$, we can also show the backgrounds in blue and green. On
the plateau we find tagged tops in roughly 20\% of the fat jets for
semileptonic top pairs and 2-4\% for $W+$jets and QCD jets.

The candidate histogram in the left panel of Fig.~\ref{fig:fat}
exceeds the numbers for top decay products in the fat jet because
there exist fat jets whose three main subjets are not from a top decay and
accidentally give $m_t$. The central panel of Fig.~\ref{fig:fat} shows
the composition of each type separately for candidates (dotted) and
tags (solid). Indeed, a considerable fraction of candidates are of
type~3. On the other hand, type~3 and type~2 tags are effectively
rejected by the mass plane cut, so most of the tagged tops are of
type~1. The fraction of type~2 and type~3 tags ranges around 2-4\%,
similar to QCD and $W+$jets backgrounds.

The right panel shows the fractions of candidate and tagged fat jets
including a hadronic top, so they are also constrained to belong to
the second category.  Compared with the central panel we are now less
likely to encounter type~2 or type~3 candidates.  Most tagged tops are
now of type~1 and most type~2 and type~3 tags correspond to fat jets
which do not include a hadronic top.  Consequently, we find that there
is not much room to improve our algorithm in selecting subjets
after applying the mass drop criterion.

\section{Alternative jet algorithms}
\label{sec:algos}

The combination of the Cambridge--Aachen clustering
algorithm~\cite{ca_algo} with a mass
drop criterion~\cite{bdrs} is a core feature of the {\sc HEPTopTagger}
and hence not negotiable. However, in the mass reconstruction after
filtering the C/A algorithm should be compared to alternative jet
algorithms, like $k_T$~\cite{kt_algo} or anti-$k_T$~\cite{anti_algo}.
After identifying three subjets based on the mass drop criteria there
are two steps left to extract the $b$, $W_1$, and $W_2$ subjets:
filtering~\cite{bdrs} and reclustering.  We find that the choice of
jet algorithm for the filtering does not have a visible effect on the
efficiency at particle level, while for the reclustering step it does.
In our explanations we therefore focus on effects of this reclustering
which combines five filtered subjets into three top decay subjets
while we always use the same jet algorithm for filtering and for
reclustering.\bigskip

Our first result is that the anti-$k_T$ algorithm fails to reliably
identify the three hard top decay products.  It tends to first
recombine a pairing with large transverse momentum, such that of the
three reclustered subjets one is very hard and two are very soft.
Applying our $W$ and top mass cuts will typically reject such
unbalanced combinations.

\begin{figure}[b]
\includegraphics[width=0.3\textwidth]{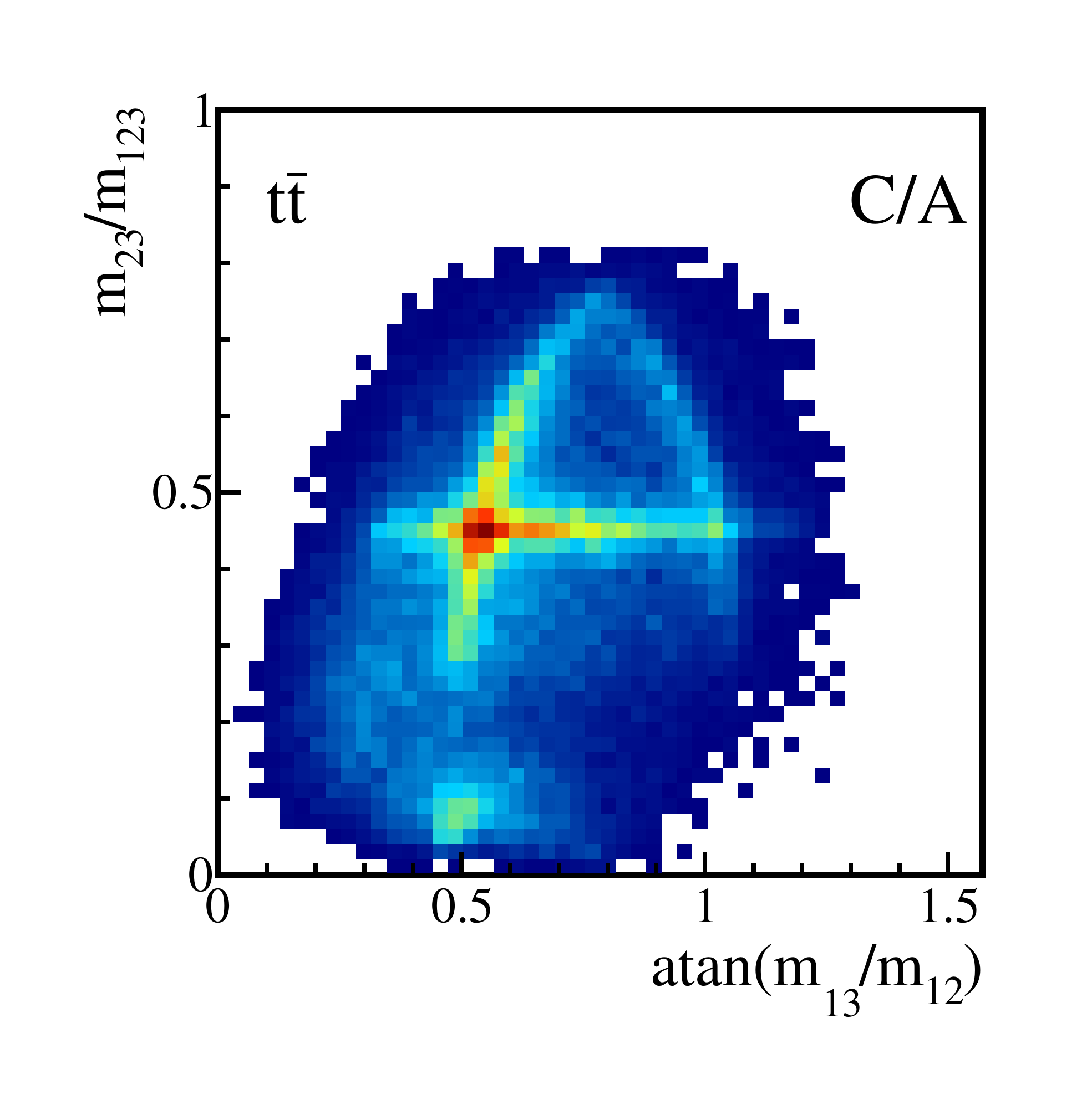}
\includegraphics[width=0.3\textwidth]{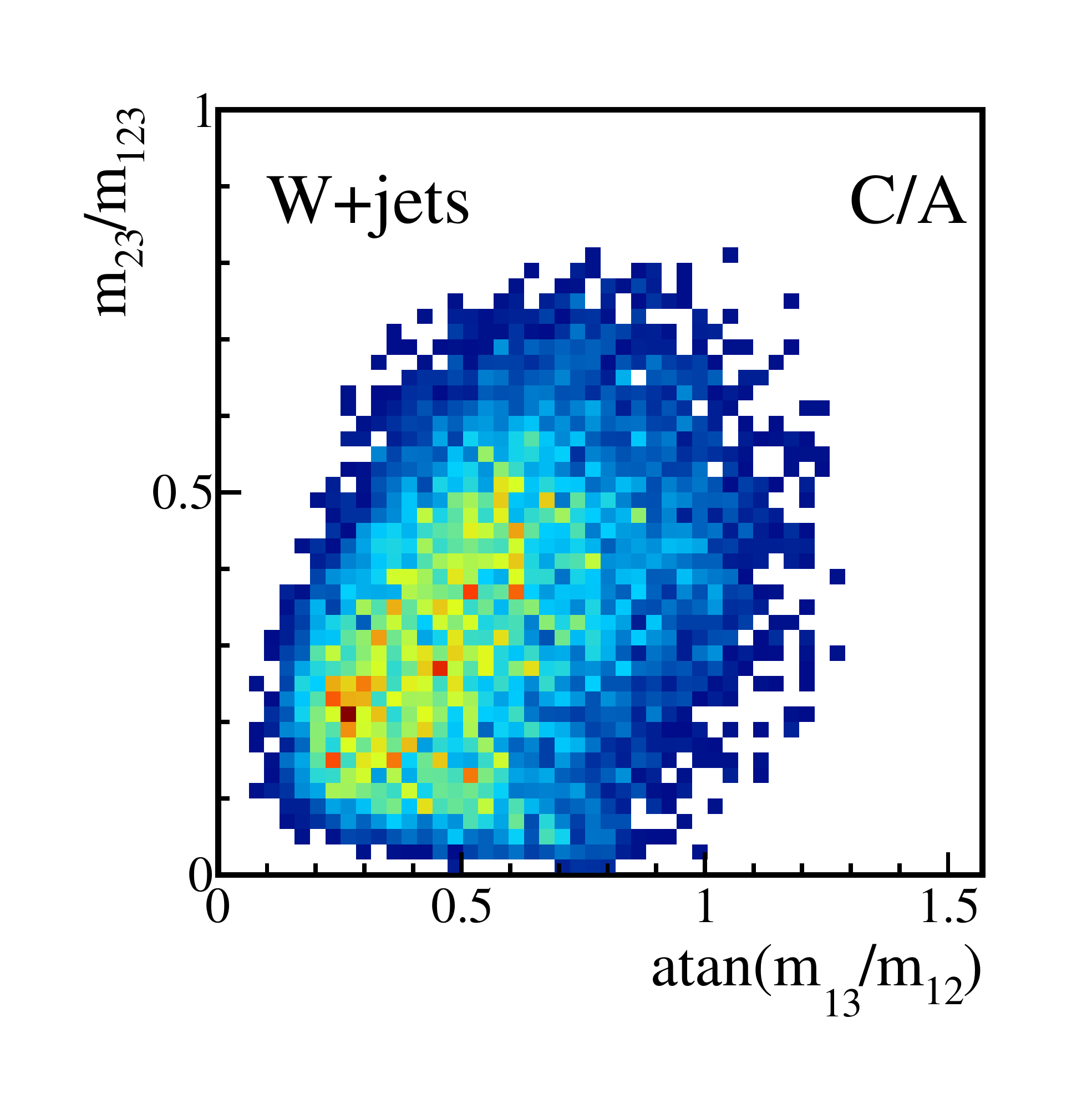}
\includegraphics[width=0.3\textwidth]{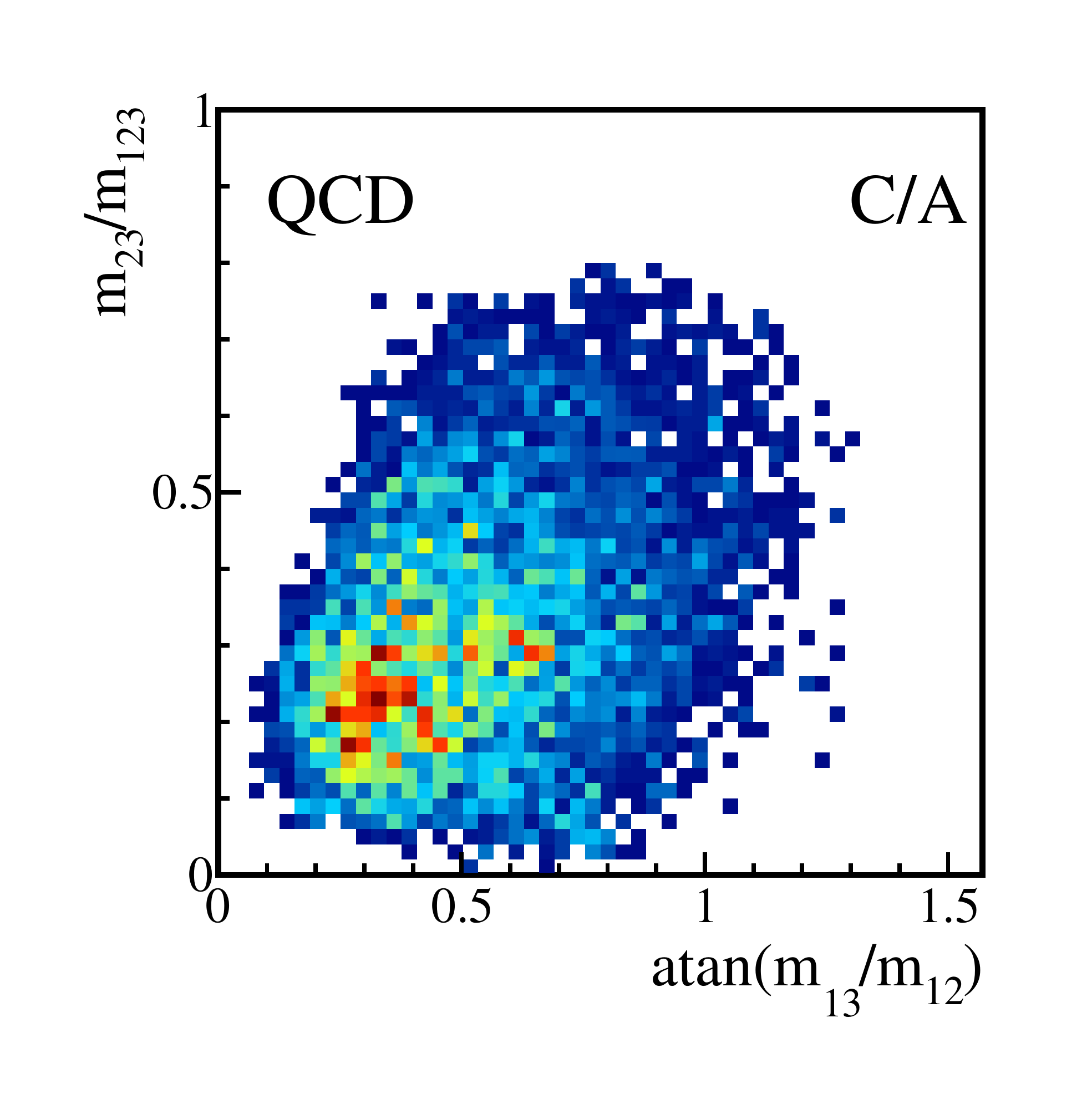} \\[-3mm]
\includegraphics[width=0.3\textwidth]{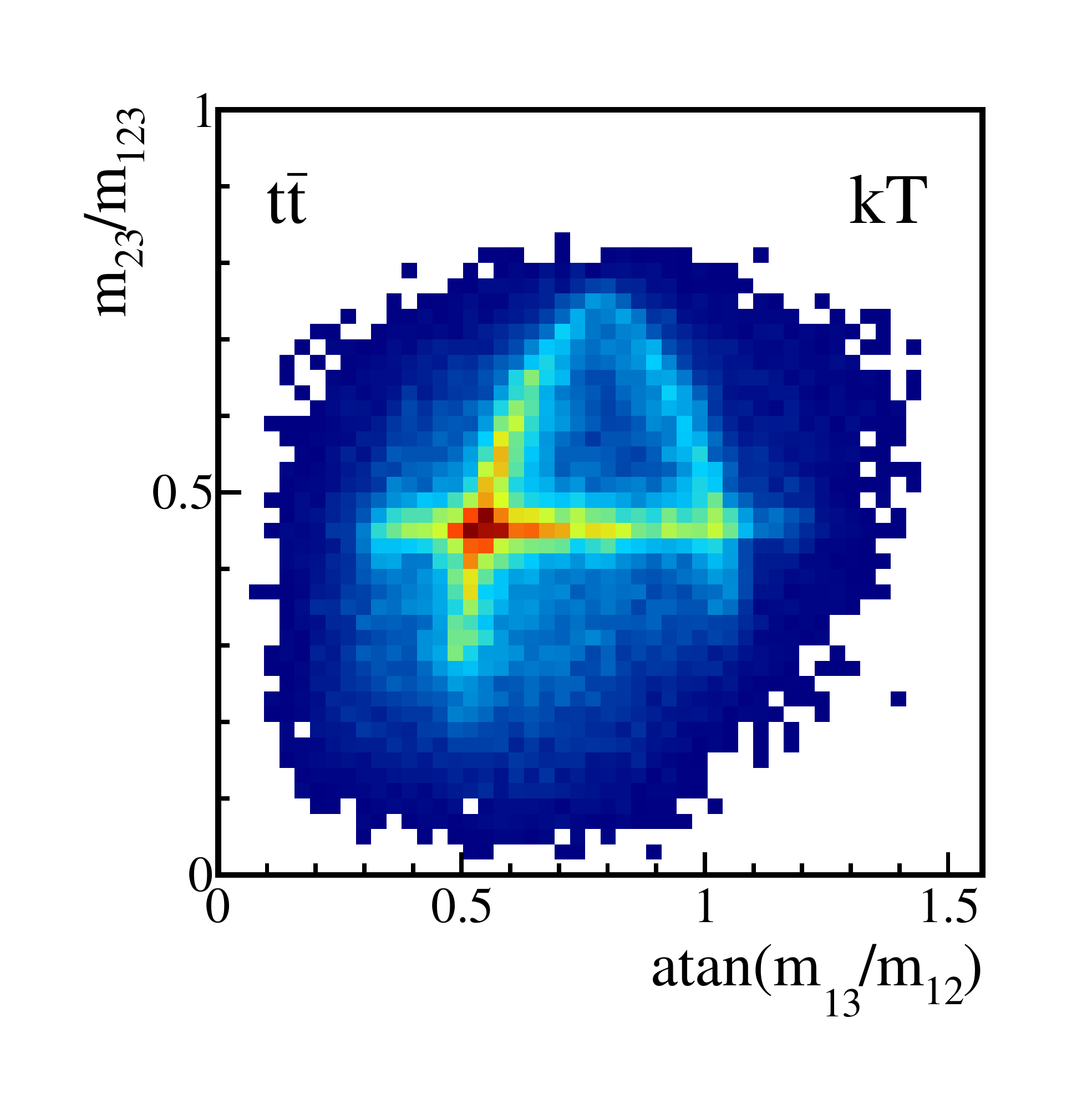}
\includegraphics[width=0.3\textwidth]{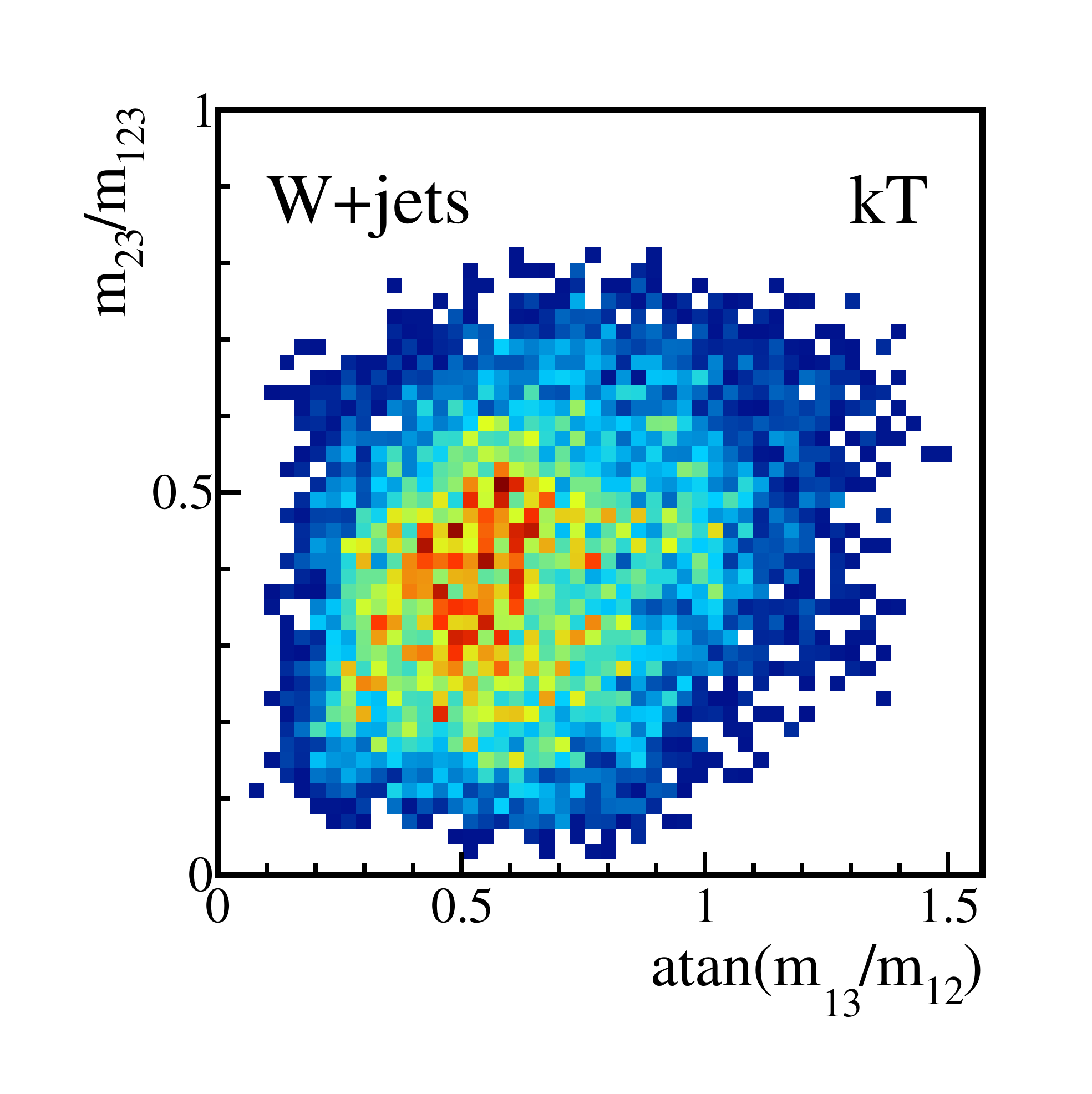}
\includegraphics[width=0.3\textwidth]{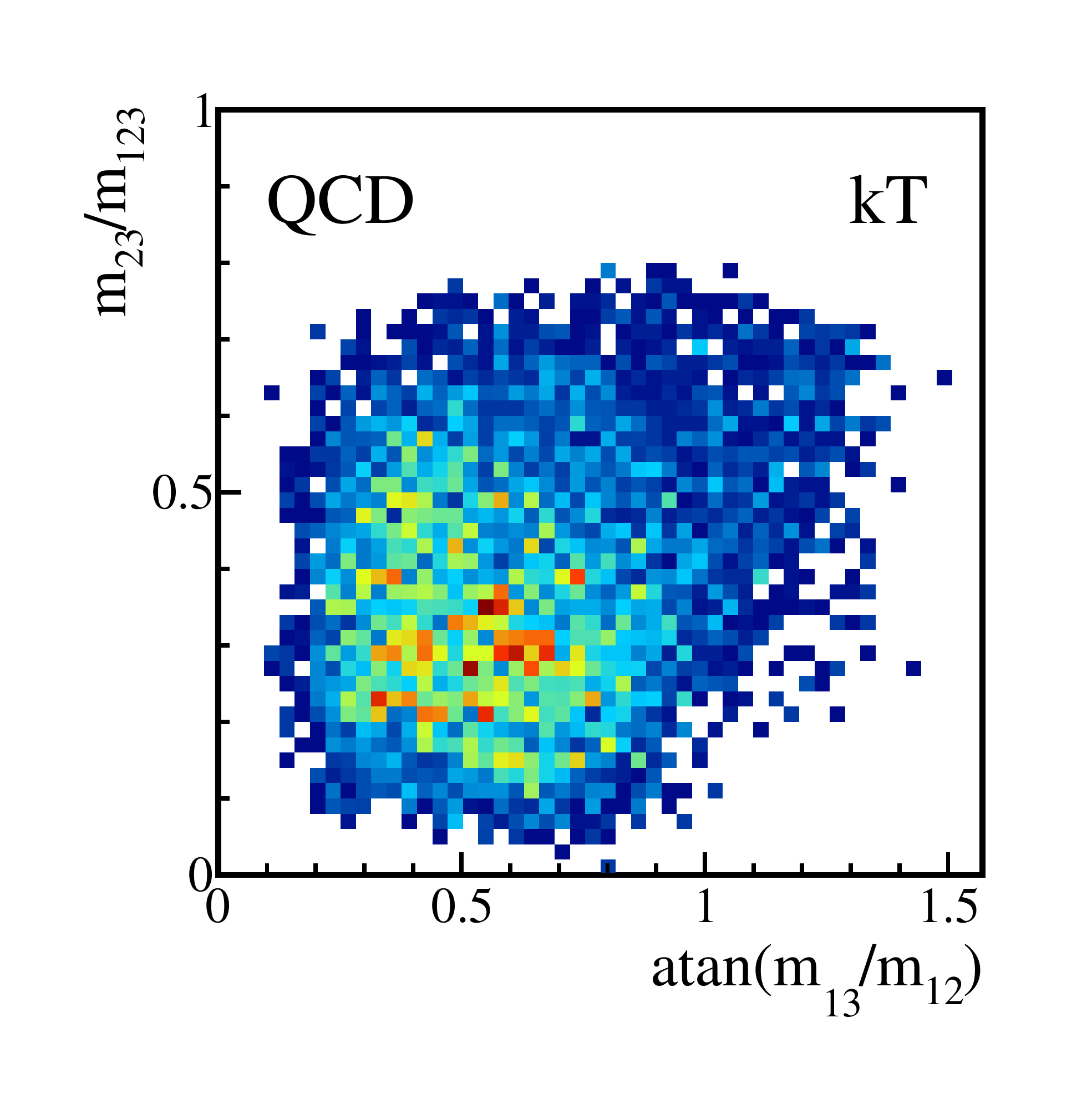}
\vspace*{-4mm}
\caption{Mass plane for top candidates in $t\bar{t}$, $W+$jets and QCD
  events (left to right).  The upper three panels use C/A reclustering
  while the lower three panels use $k_T$ reclustering.}
 \label{fig:2d_all}
\end{figure}

Using the C/A algorithm a similar problem arises, but only for very
large $p_{T,t}$, where the five filtered subjets are close.  According
to QCD the two softest filtered subjets should then be merged into the
main three subjets, which the C/A algorithm achieves as long as the
three main subjets are geometrically well separated. Once the
geometric distances become small the probability to correctly
reconstruct the three main subjets decreases.  Such signal events
appear in the lower left corner of the ($\arctan m_{13}/m_{12}$
vs. $m_{23}/m_{123}$) distributions~\cite{heptop} shown in
Fig.~\ref{fig:2d_all}.  Events close to $x$-axis have $m_{23}\sim 0$
which usually means $p_3 \sim 0$ for the third-hardest decay subjet.\bigskip

\begin{figure}[t]
\includegraphics[width=0.32\textwidth]{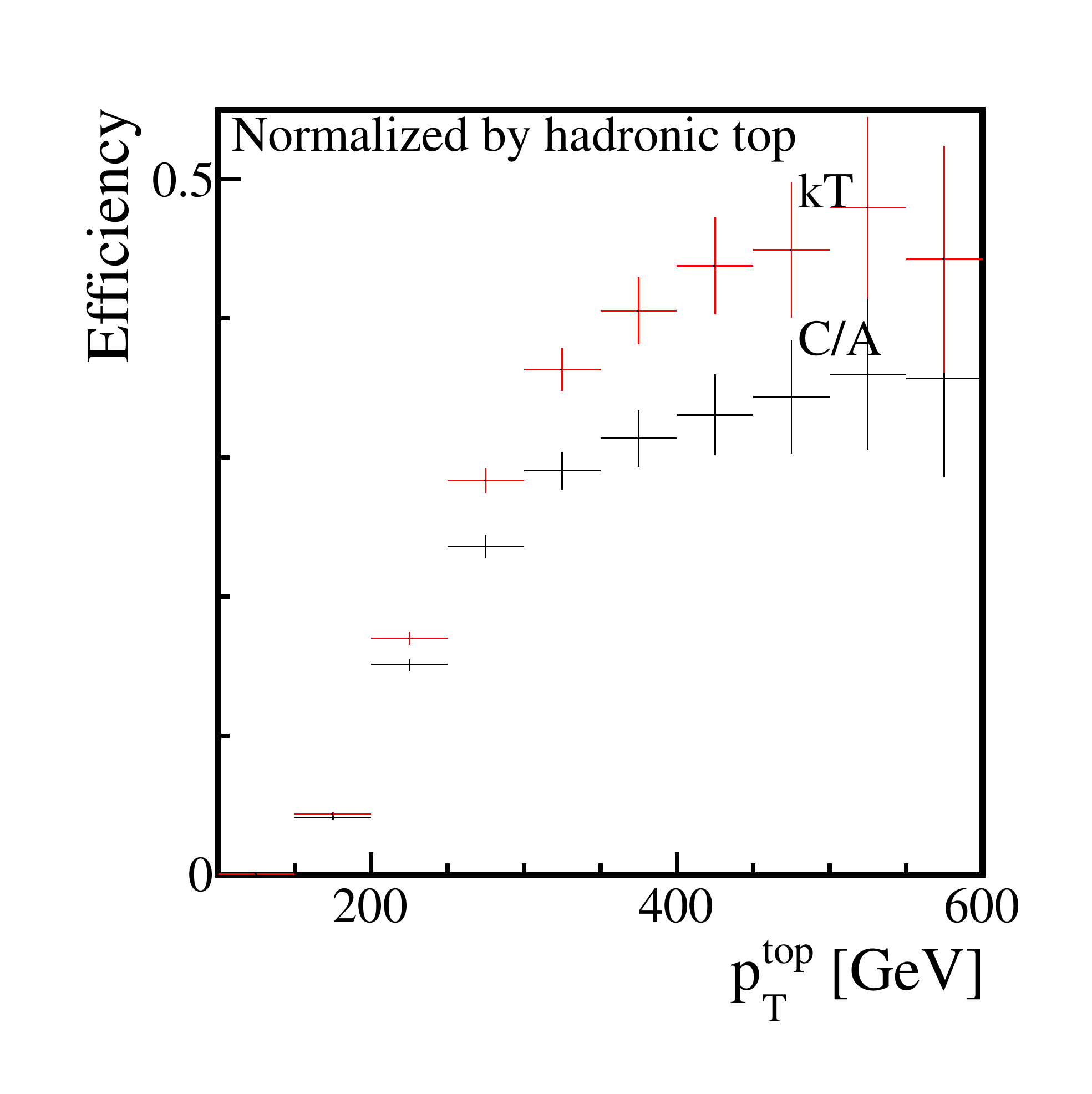}
\includegraphics[width=0.32\textwidth]{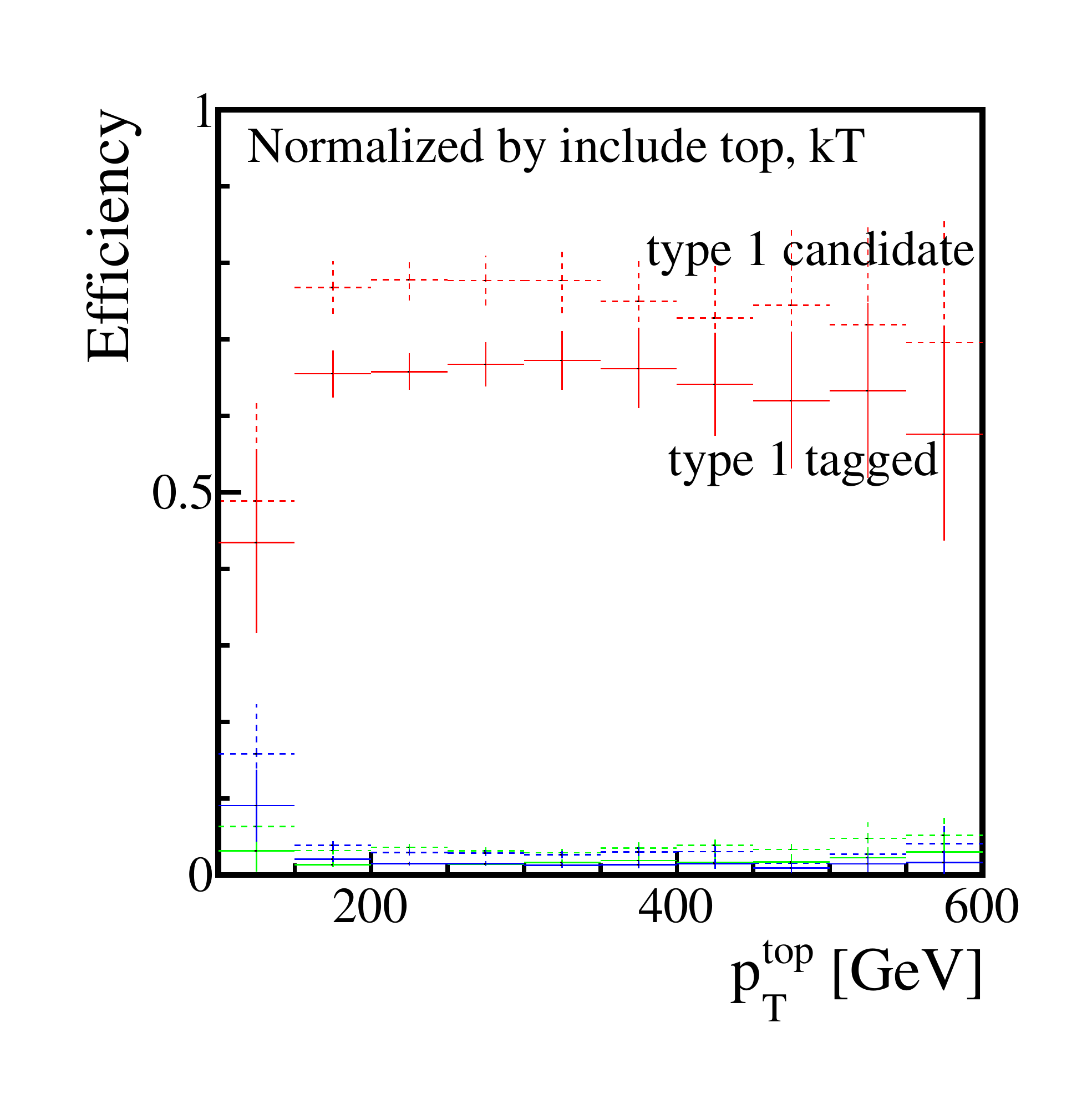}
\includegraphics[width=0.32\textwidth]{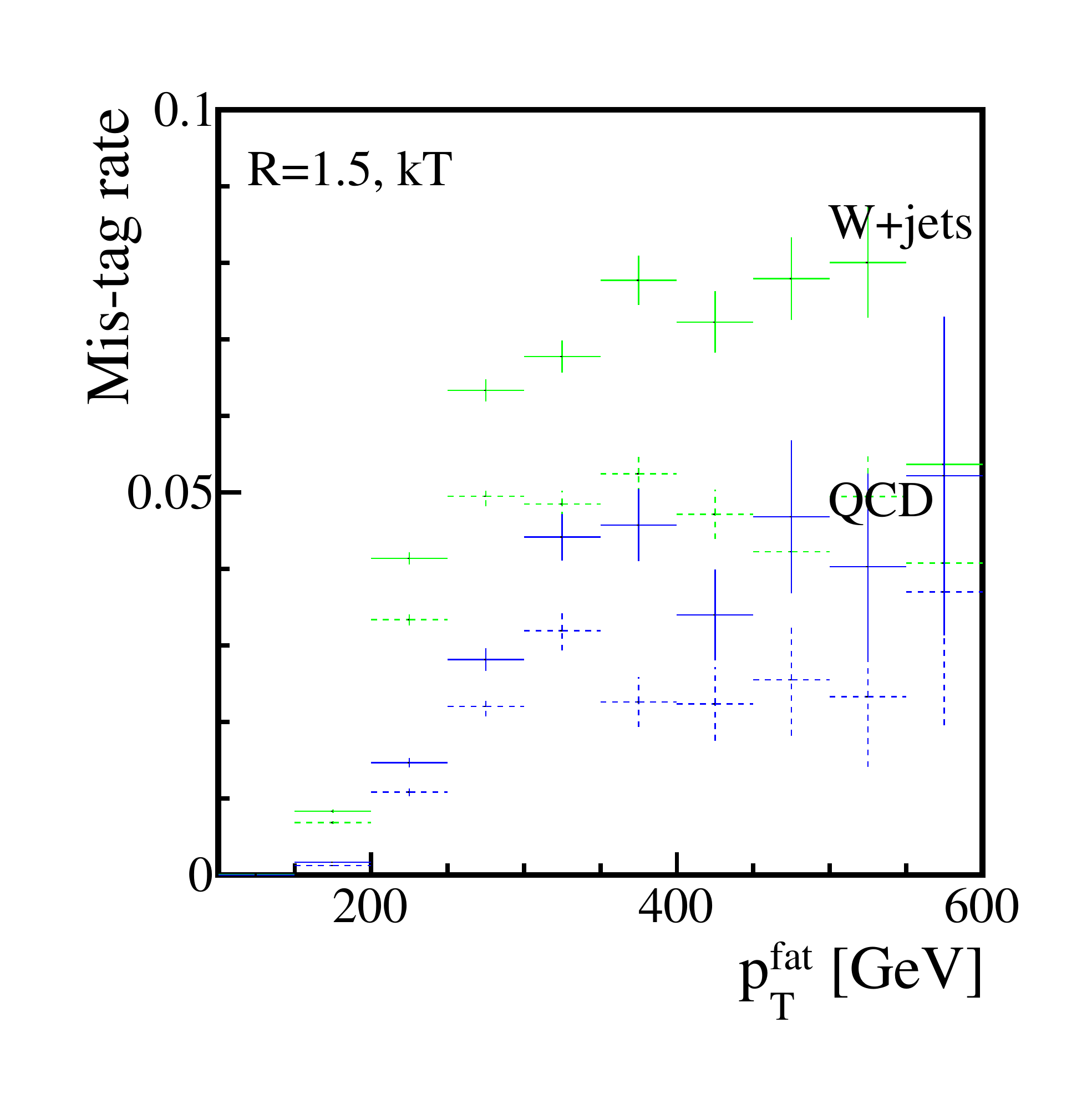}
\vspace*{-4mm}
\caption{Left: signal efficiencies relative to the number of hadronic
  tops for the $k_T$ (red), C/A (black) and anti-$k_T$ (with C/A
  clustering, blue) algorithms. Center: $k_T$ tagging efficiency
  relative to the number of tops included in a fat jet for type~1
  (red), type~2 (green), and type~3 (blue) tags.  Right: mis-tag rates
  for QCD (blue) and $W+$jets (green) for the C/A (dotted) and $k_T$
  algorithm (solid).}
\label{fig:5}
\end{figure}

The $k_T$ algorithm recombines soft filtered subjets most reliably, so
it can resolve the three main top decay products best.  We can see
this in the lower panels of Fig.~\ref{fig:2d_all} where hardly any
signal events migrate to $m_{23}\sim 0$.  As a result, the signal
efficiency of passing the mass plane cut increases. This tagging
efficiency we show in Fig.~\ref{fig:5}.  For both cases the filtering
and reclustering algorithms are the same.  In the low-$p_T$ regime the
difference between the two algorithm is indeed small.  In the central
panel we see that unlike the C/A results in Fig.~\ref{fig:had} the
$k_T$ efficiency hardly decreases towards high $p_{T,t}$, \ie the
efficiency of the mass plane cut is essentially constant.  The
fractions of type~2 and type~3 tags which include a top also decrease
because the subjet momentum reconstruction improves. If the subjets
are better matched to the hard top decay products more tagged tops are
categorized as type~1.\bigskip

Going beyond the signal, the same feature of fewer events with
$m_{23}\sim 0$ also appears for the backgrounds.  The mass plane cut
then leads to a less efficient rejection in particular for soft
masses. This increase of mis-tag probabilities we also show in
Fig.~\ref{fig:5}. The solid crosses show the results for the $k_T$
algorithm while the dotted crosses show those for the C/A algorithm.
Quantitatively, the mis-tag efficiencies increase by a slightly
larger factor than the tagging efficiency, so switching to the $k_T$
algorithm does not improve $S/B$ but can improve $S/\sqrt{B}$
slightly. In addition, these results are obtained without detector
simulation and pileup. Switching to the
$k_T$ algorithm might have additional implications from both of them,
so only a detailed experimental analysis can determine which jet algorithm to
use in the {\sc HEPTopTagger} framework, and its result might well be process
dependent.

\section{Pruning}
\label{sec:prune}

From the Higgs tagger based on the C/A algorithm and a mass drop
criterion~\cite{bdrs} we know that it can be advantageous to combine
filtering and pruning in the tagging procedure~\cite{soper_spanno}.
Pruning removes soft radiation while clustering the fat
jet~\cite{pruning, scetfact}: first, a sequential jet
algorithm combines unfiltered subjets until no pair of constituents is
geometrically closer than $R_\text{cut}$, representing an effective
subjet cone size usually associated with an intrinsic fatjet scale. We choose
$R_\text{cut} = m_j/ p_{T,j}$. This cutoff can act differently 
for different underlying jet algorithms. After this recombination
the unfiltered subjet merging continues, but with the additional restriction
that each combined pair of subjets has to be sufficiently hard,
\begin{equation}
z=\frac{\text{min} (p_{T,i},p_{T,j})}{|\vec{p}_{T,i} + \vec{p}_{T,j}|}>z_\text{cut} \; ,
\end{equation}
where we choose $z_\text{cut} = 0.1$. Otherwise, the constituents $i$
and $j$ are not combined and the one with smaller transverse momentum
is discarded. This algorithm continues until all constituents have 
been combined or eliminated.

There are two ways we can include pruning in our top tagger.  First,
we can prune the fat jet before we run the C/A algorithm extracting
the relevant splittings using the mass drop criterion.  Alternatively,
we can use the pruning procedure in parallel to filtering procedure and combine the two pieces of information.  This approach ensures
that the additional pruning step does not affect the performance of
the rest of the tagging algorithm, so we investigate it in this
section.\bigskip

\begin{figure}[t]
\includegraphics[width=0.32\textwidth]{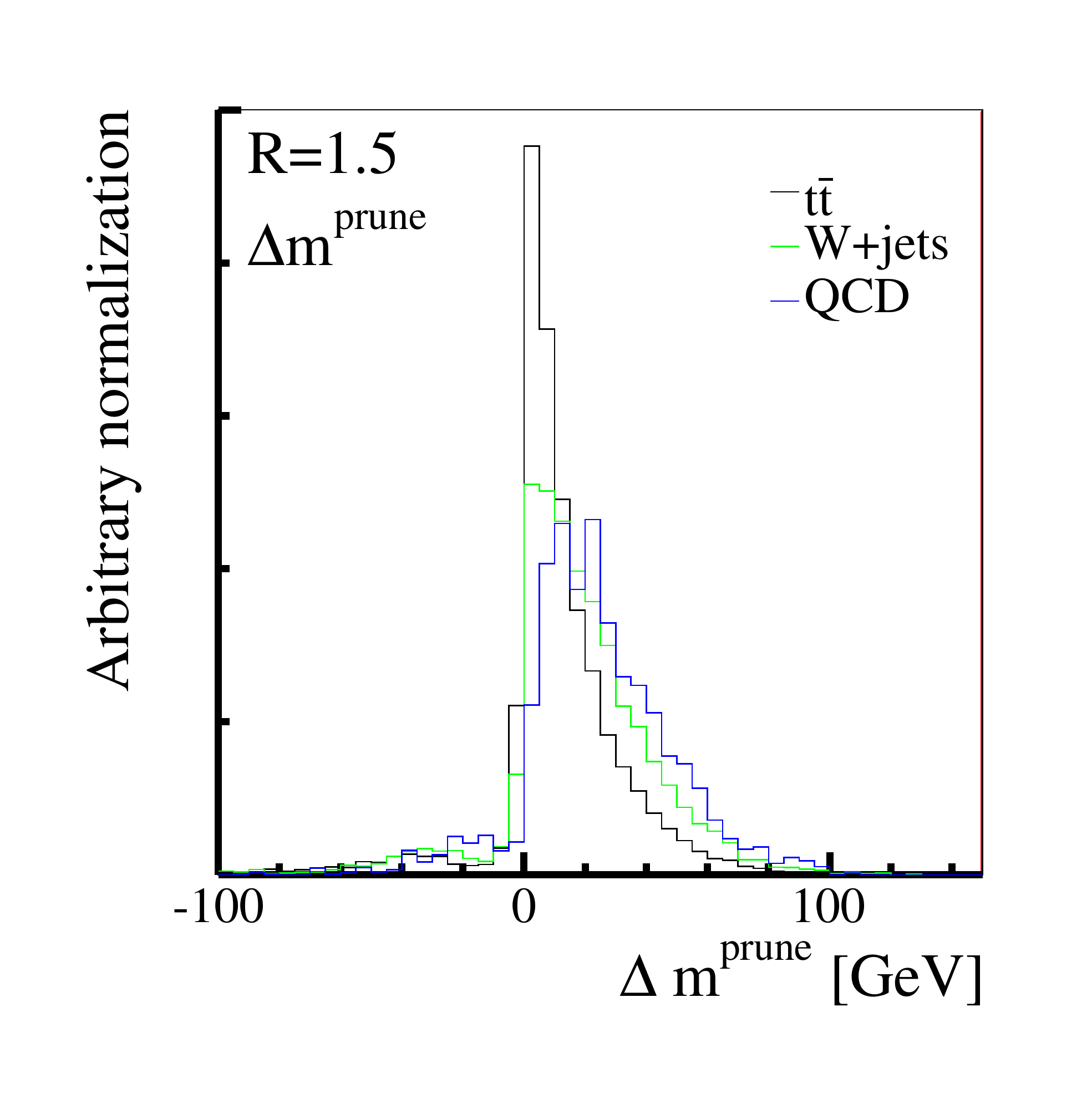}
\includegraphics[width=0.32\textwidth]{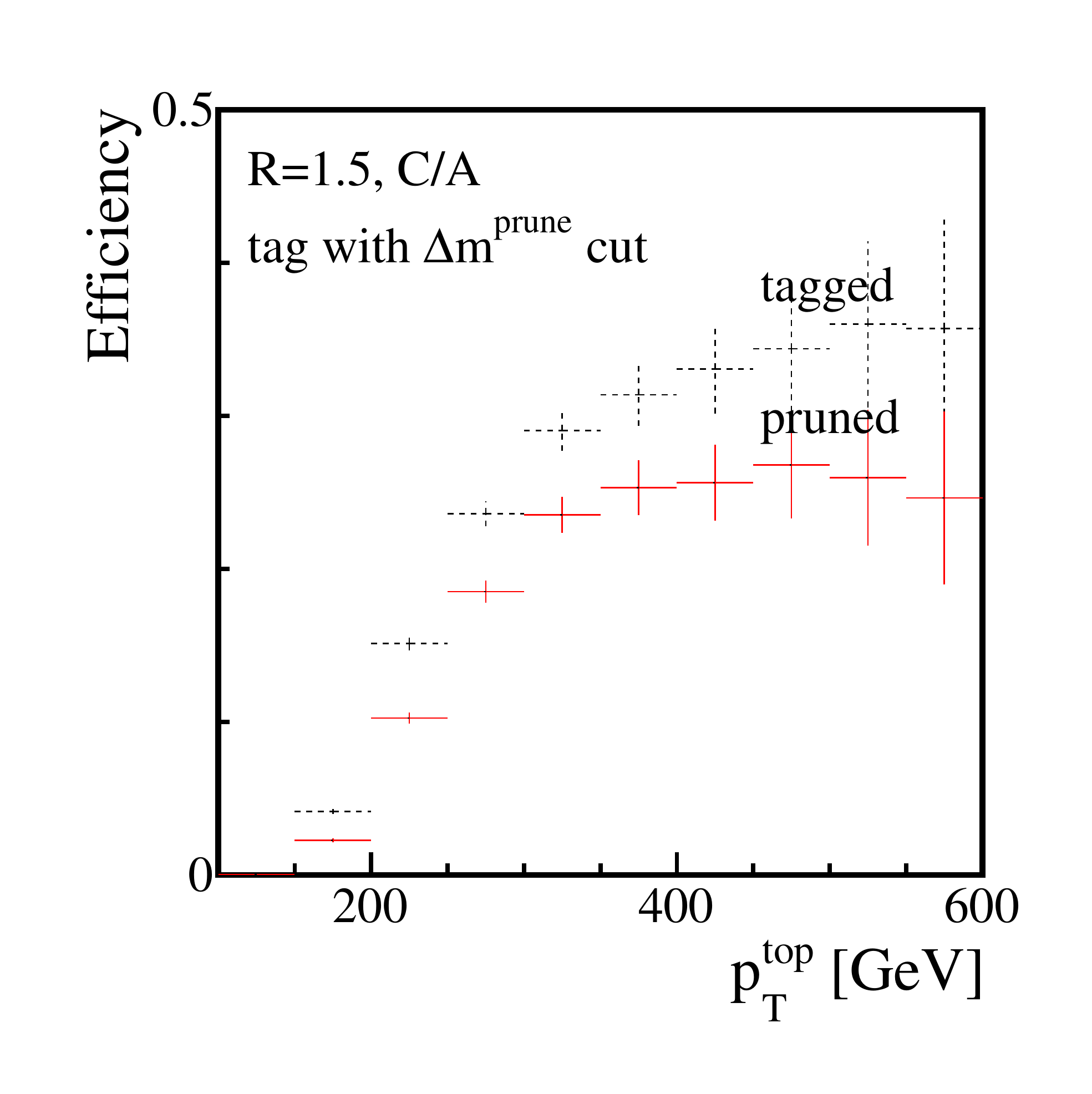}
\includegraphics[width=0.32\textwidth]{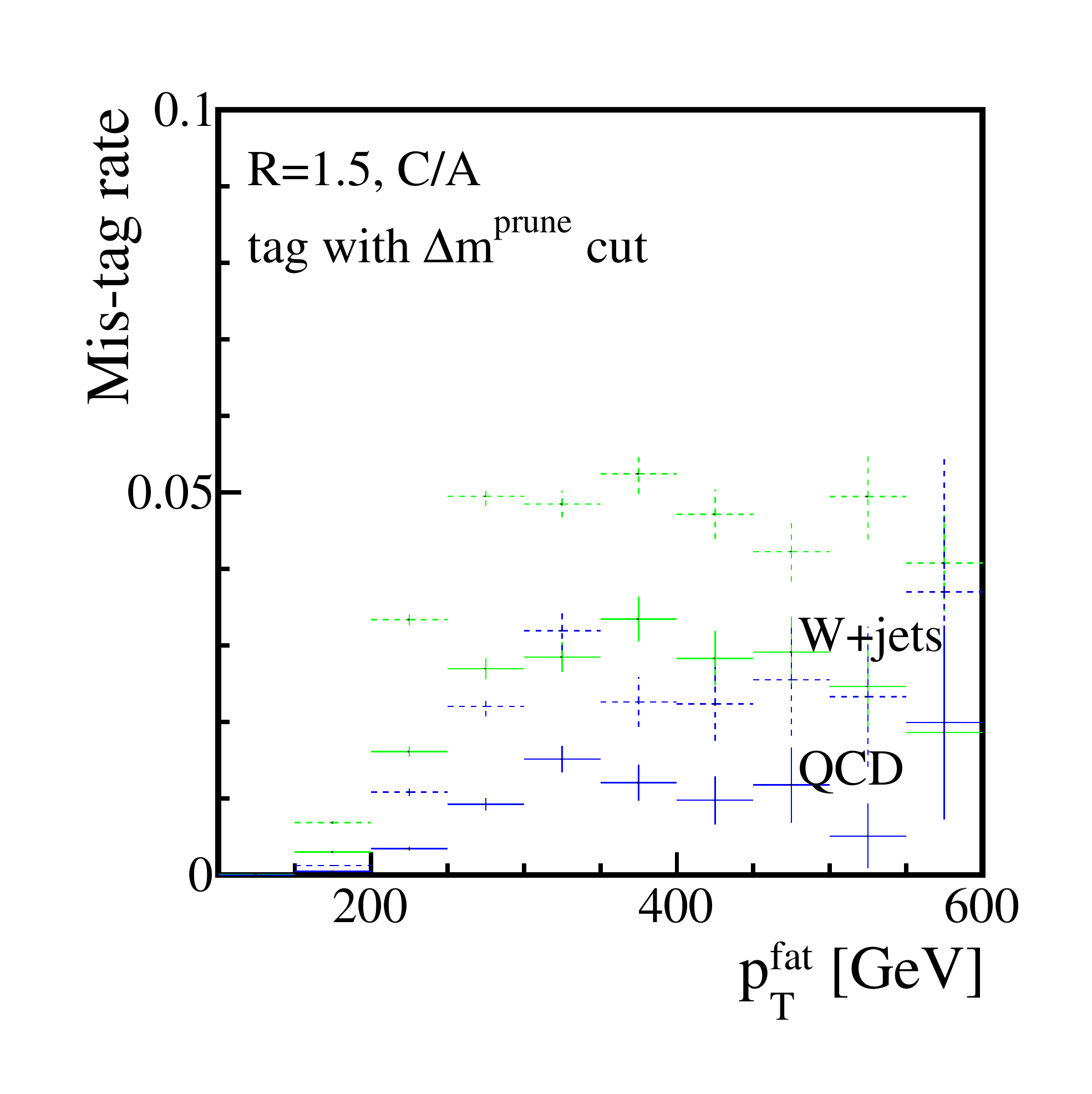}
\vspace*{-4mm}
\caption{ Left: $\Delta m^\text{prune}$ distribution for type~1 tags
  in $t\bar{t}$ (black), $W$+jets (green), and QCD (blue) events.
  Center: signal efficiencies as a function of $p_{T,t}$ without
  (black) and with (red) the pruning cut $\Delta m^\text{prune} < 20$GeV defined in Eq.\eqref{eq:prunecut}.
  Right: mis-tag probabilities for QCD (blue) and $W+$jets (green) as
  functions $p_T^\text{fat}$ with (solid) and without (dotted) the
  pruning cut.}
\label{fig:6}
\end{figure}

\begin{table}[b]
\begin{tabular}{c||r|rrr|rrr}
\hline
&\ \ \ tagged & $\Delta m^\text{prune}< 15 $ & $20$ & $30$ & 
                $\Delta m^\text{unfilter}<15 $ & $20$ & $30$ \cr
\hline
$t\bar{t}$ [fb] &
14156& 7773 (55\%)& 9072 (64\%)& 10875 (77\%)& 6237 (44\%)& 7926 (56\%)& 10505 (74\%)\cr
\hline
type~1 & 10318& 6255 (61\%)& 7152 (69\%)& 8316 (81\%)& 5141 (50\%)& 6377 (62\%)& 8129 (79\%)\cr
type~2 & 1336& 551 (41\%)& 693 (52\%)& 893 (67\%)& 403 (30\%)& 570 (43\%)& 847 (63\%)\cr
type~3 & 2503& 967 (39\%)& 1227 (49\%)& 1666 (67\%)& 693 (28\%)& 979 (39\%)& 1529 (61\%)\cr
\hline
$W$+jet [fb] &
6590  & 2716  (41\%)& 3373 (51\%)& 4459 (68\%)& 2052 (31\%)& 2797 (42\%)& 4162 (63\%)\cr
$\epsilon_{S/B}$          &1&  1.33	&	1.25	&	1.14	&	1.41	&	1.32	&	1.18\cr
$\epsilon_{S/\sqrt{B}}$&1& 0.86	&	0.9	&	0.93	&	0.79	&	0.86	&	0.93\cr
\hline
QCD [pb] &
1229 & 359 (29\%) & 474 (39\%) & 719 (59\%) & 207 (17\%)& 331 (27\%) & 609 (50\%)\cr
$\epsilon_{S/B}$           &1 &  1.88	&	1.66	&	1.31	&	2.62	&	2.08	&	1.5\cr
$\epsilon_{S/\sqrt{B}}$&1 & 1.02	&	1.03	&	1	&	1.07	&	1.08	&	1.05 \cr
\hline
\end{tabular}
\caption{Tagged top rates (in fb for $t\bar{t}$ and $W$+jets and pb
  for QCD jets) after cuts on $\Delta m^\text{prune}$ or $\Delta
  m^\text{unfilter}$.  The percentages are relative to the numbers
  without pruned or unfiltered mass cut in each category.
  $\epsilon_{S/B}$, $\epsilon_{S/\sqrt{B}}$ denote improvement factors
  relative to no cuts.
  }
\label{tab:prune}
\end{table}

\begin{figure}[t]
\includegraphics[width=0.32\textwidth]{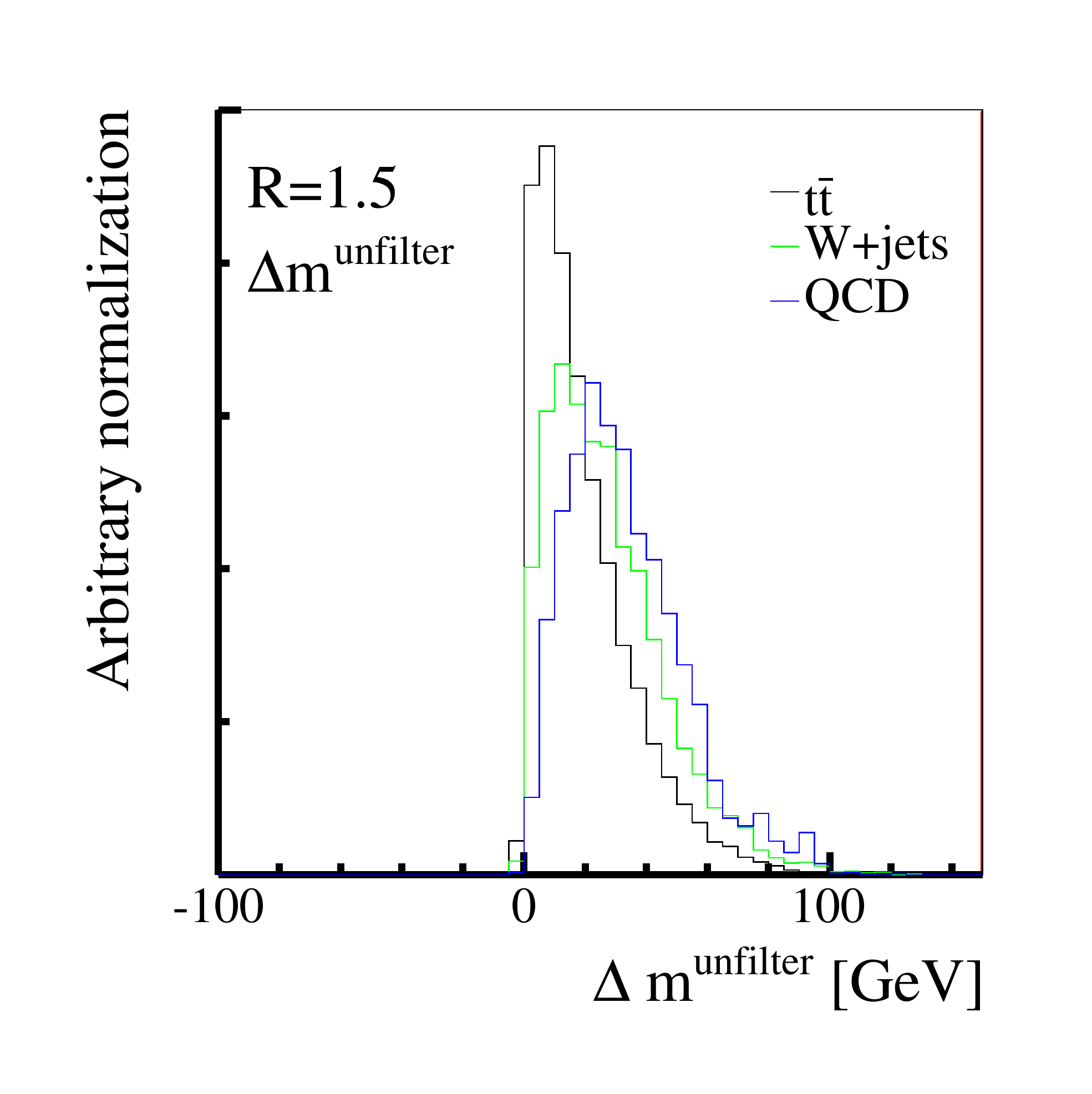}
\includegraphics[width=0.32\textwidth]{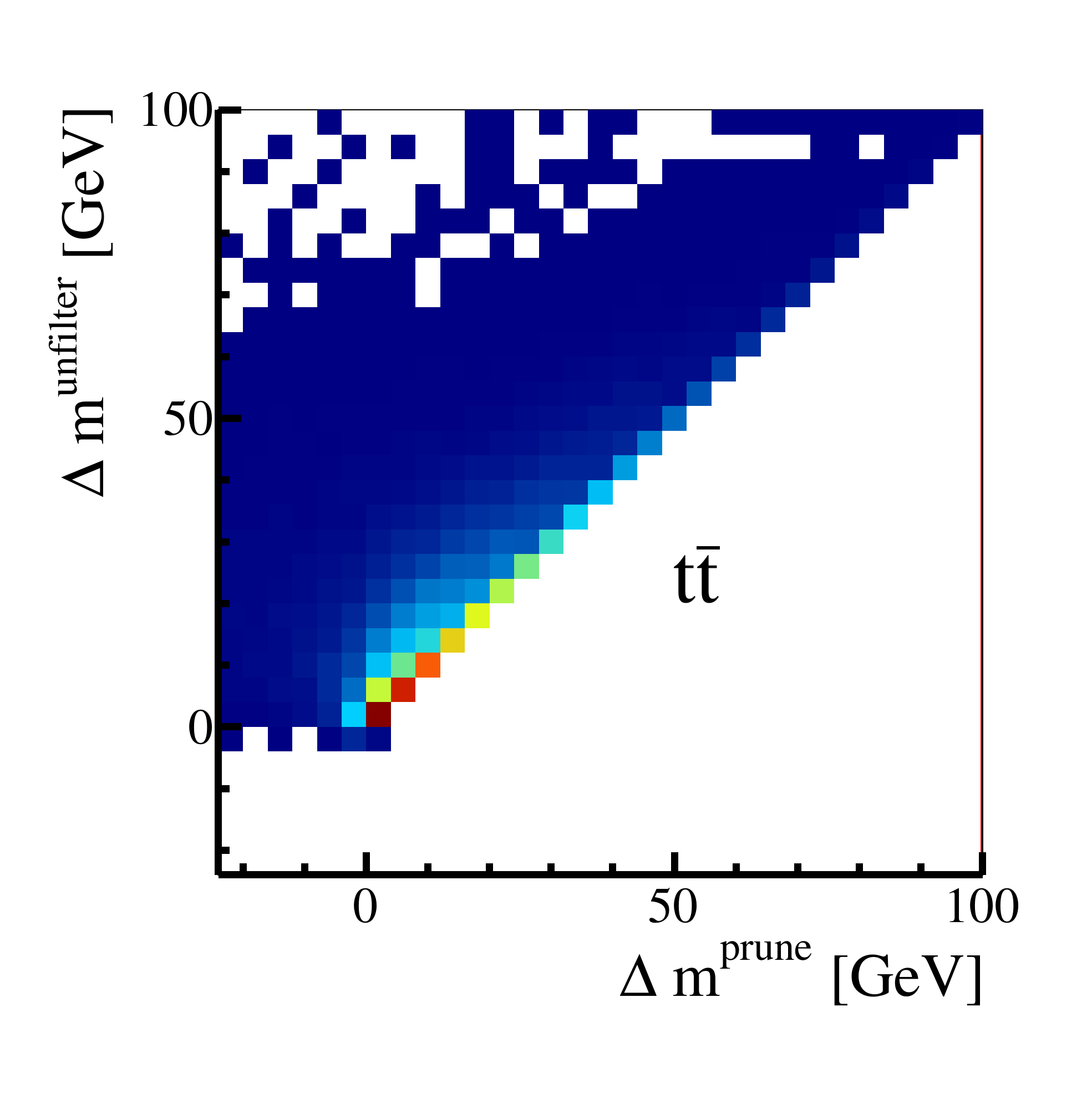}
\includegraphics[width=0.32\textwidth]{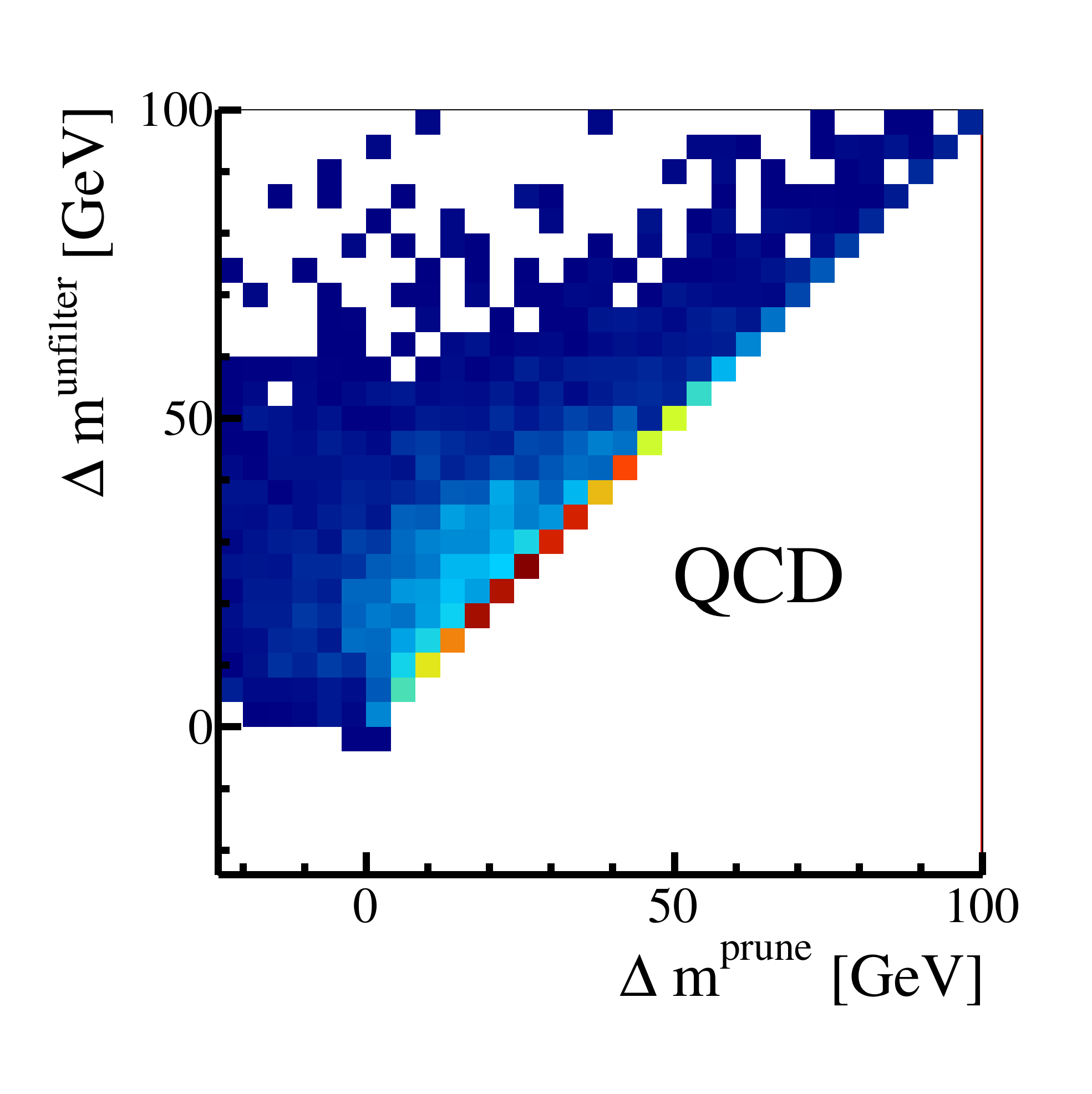}
\vspace*{-4mm}
\caption{Left: $\Delta m^\text{unfilter}$ for type~1 $t\bar{t}$ (black),
  $W$+jets(green), and QCD (blue) events. Center and right:
  two-dimensional correlation of the two mass differences.}
\label{fig:7}
\end{figure}

Since pruning is originally targeted at removing soft radiation, its
impact is similarly to that of filtering~\cite{bdrs}.  To quantify the
difference we apply pruning to the constituents of the three subjets
which we obtain after the usual mass drop criterion. These three
subjets are selected such that they give the best filtered top mass
among all combinations.  The difference between the two algorithms is
illustrated by the variable
\begin{equation}
\Delta m^\text{prune} = m^\text{prune} - m^\text{filter} \; ,
\end{equation}
where $m^\text{filter}$ is the filtered mass for the selected three
subjets, and $m^\text{prune}$ is the jet mass of the pruned jet.  The
left panel in Fig.~\ref{fig:6} shows $\Delta m^\text{prune}$ for
tagged tops in $t\bar{t}$ (type~1), QCD jets, and $W+$jets events. We
find that $\Delta m^\text{prune}$ is larger for background events than
for signal events.
Pruning generally collects more constituents than filtering, which
discards some of the filtered subjets, so the pruned mass increases
in a busy jet environment. Background events rely on such busy events
to mimic the generically hard top decay products.  Therefore,
selecting tags with small $\Delta m^\text{prune}$ effectively rejects
the backgrounds.  Even though we do not show them, type~2 and type~3
tags behave similar to the backgrounds samples, because QCD jet
radiation partly contributes to these tags.  Thus, pruning also
purifies the type~1 fraction for all tagged tops.\bigskip

The event numbers after imposing 
\begin{equation}
-10~\text{GeV} < \Delta m^\text{prune} <  \{15,20,30\}~\text{GeV}
\label{eq:prunecut}
\end{equation}
we show in the left half of Tab.~\ref{tab:prune}.  The percentages are
relative to the number of tagged tops in each row. The different
efficiencies we can translate into improvement factors for $S/B$ and
for $S/\sqrt{B}$. For example, we can improve $S/B$ by roughly a
factor two without loss of $S/\sqrt{B}$. In addition, the quality of
the momentum reconstruction improves with the increased fraction of type~1
tags.

This additional cut becomes more important for larger thresholds of
the subjet mass in the top tagger.  For example, compared with the
default choice $m_\text{subjet} > 50$~GeV a reduction to 30~GeV makes
the cut on $\Delta m^\text{prune}$ less efficient.  This arises
because with a larger $m_\text{subjet}$ threshold more constituents
can contribute in the background case.\bigskip

According to the above discussion the difference in the pruned mass
distribution for signal and backgrounds is due to the subjet
multiplicity inside the fat jet.  A simpler intuitive measure for
this feature is the fat jet mass before filtering. Algorithmically, we
select the relevant three subjets only after filtering, but the
original mass of the unfiltered subjets includes additional
information:
\begin{equation}
\Delta m^\text{unfilter} = m^\text{unfilter} - m^\text{filter} \; .
\end{equation}
In Fig.~\ref{fig:7} we show this distribution for signal and
backgrounds.  Indeed, the left panel is very similar to the $\Delta
m^\text{prune}$ distribution of Fig.~\ref{fig:6}. The two-dimensional
correlation confirms that almost all events for signal and background
lie on the central diagonal of the $\Delta m^\text{prune}$ vs $\Delta
m^\text{unfilter}$ plane.  

In the right half of Tab.~\ref{tab:prune}
we show the corresponding efficiencies after cutting on $\Delta
m^\text{unfilter}$. From both variables we can obtain significant
improvements on $S/B$, and it remains an experimental question which
of them is more stable once we include detector effects and pile-up.

\section{Fatter jets}
\label{sec:size}

In the Standard Model the cross section for top pairs falls very
steeply with increasing transverse momentum. The fraction of top pair
event above different $p_{T,t}^\text{min}$ values at a 14~TeV LHC we calculated using {\sc Alpgen}~\cite{alpgen}:
\begin{center} \begin{tabular}{l|rrrrrrrr}
\hline
$p_{T,t}^\text{min}~[\gev]$ &   0   & 100 & 150 & 200 & 250 & 300 & 400 & 500 \cr
\hline 
fraction &100\% &53\%  &28\%  & 14\% & 6.8\%  & 3.4\% &  0.96\% & 0.33\% \cr
\hline
\end{tabular} \end{center}
Extending the top tagging reach by 50~GeV towards smaller $p_{T,t}$ 
corresponds to doubling the number of accessible top pairs.\bigskip

The question becomes where the observed limitations of top tagging in
this regime really come from and whether these constraints can be
removed.  From the left panel of Fig.~\ref{fig:had} we know that the
fraction of hadronic tops which can be included in a fat jet rapidly
drops around $p_{T,t} = 200 - 250$~GeV.
Compared to a well suited data sample with $p_{T,t}>300$~GeV the
tagging probability roughly drops to half its value.  On the other
hand, in the right panel of Fig.~\ref{fig:had} we see how the fraction
of tagged tops relative to the number of hadronic tops included in a
fat jet increases. This suggests that larger values than $R=1.5$ should
significantly improve the tagging efficiency around $200 -
250$~GeV. In this section we will study an increase to $R=1.8$ for our
standard {\sc HEPTopTagger} setup to test such an option.\bigskip

\begin{figure}[t]
\includegraphics[width=0.3\textwidth]{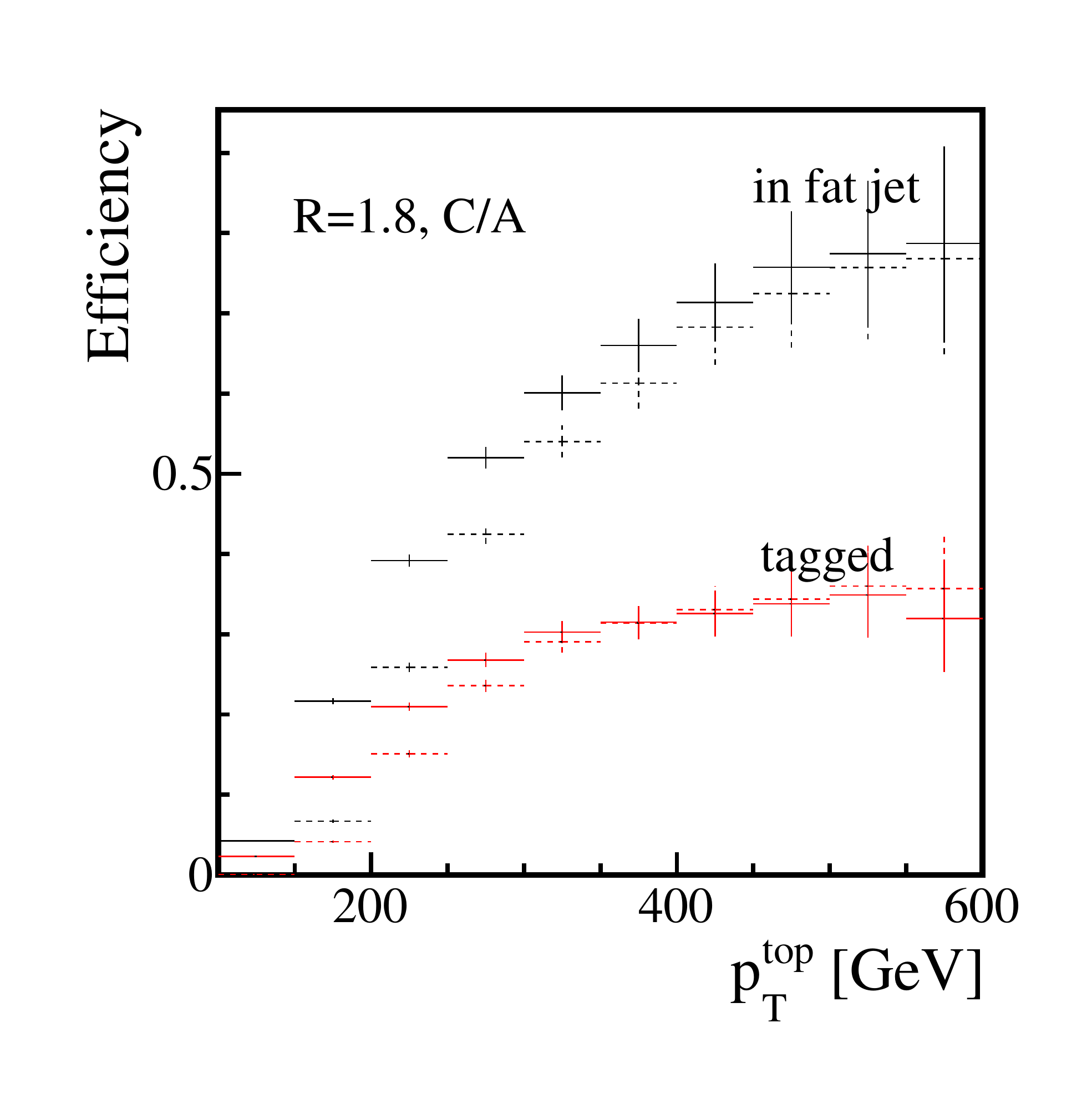}
\includegraphics[width=0.3\textwidth]{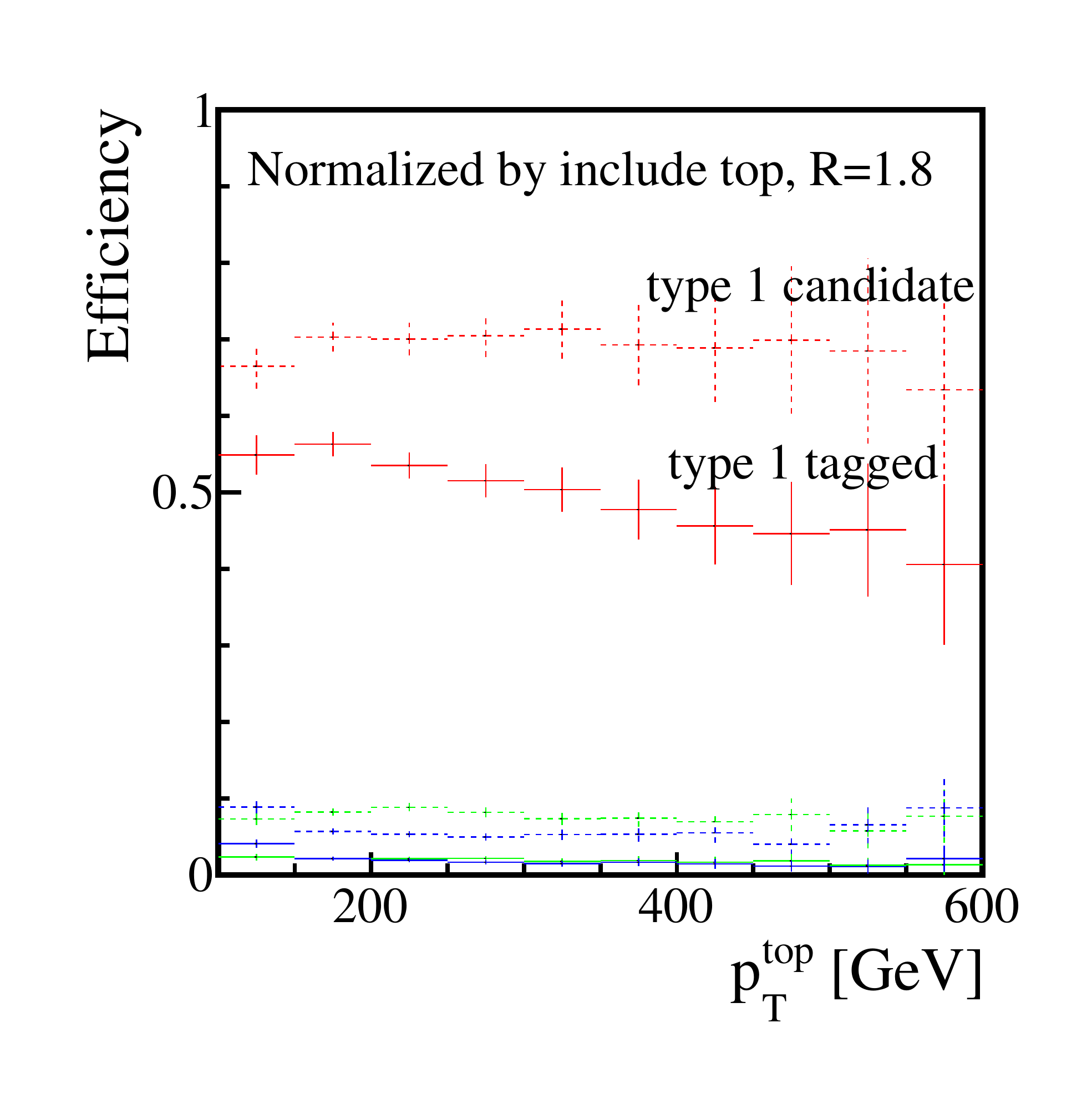}
\includegraphics[width=0.3\textwidth]{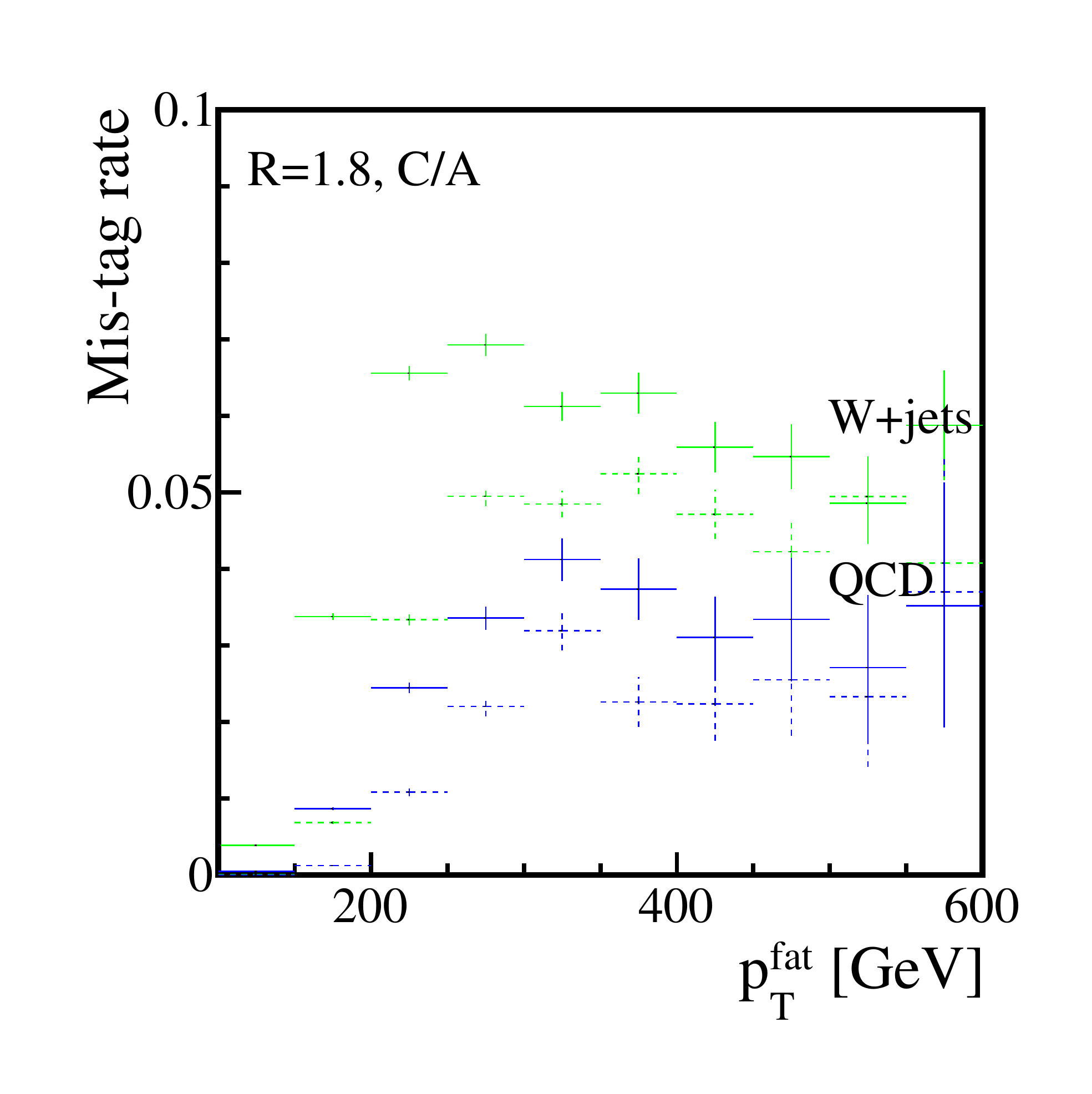}
\vspace*{-4mm}
\caption{ Left: tagging efficiency (solid red) and the fraction of
  tops included in the fat jet for $R=1.8$ (solid) and for $R=1.5$
  (dotted) as a function of $p_{T,t}$ for semileptonic $t\bar{t}$
  events.  Center: tagging candidates and tags relative to all tops
  included in a fat jet. Type~2 and type~3 tags are shown in green and
  blue. Right: mis-tag rate as a function of fat jet $p_T$ for QCD
  (blue) and $W+$jets (green).}
\label{fig:18}
\end{figure}

\begin{table}[b]
\begin{tabular}{c||r|rrr|rrr}
\hline
$R=1.8$ &\ \ \ tagged [fb] & $ \Delta m^\text{prune}<15 $ & $20$ & $30$ & $\Delta 
m^\text{unfilter}<15 $ & $20$ & $30$ \cr
\hline
$t\bar{t}$ [fb] & 27853& 10695(38\%)& 13221(47\%)& 17453(63\%)& 8253(30\%)& 11131(40\%)& 16148(58\%)\cr
\hline
type~1 &17502& 8114(46\%)& 9716(56\%)& 12195(70\%)& 6463(37\%)& 8403(48\%)& 11507(66\%)\cr
type~2 & 3628& 934(26\%)& 1252(35\%)& 1847(51\%)& 655(18\%)& 996(27\%)& 1666(46\%)\cr
type~3 & 6723& 1647(24\%)& 2252(34\%)& 3410(51\%)& 1135(17\%)& 1732(26\%)& 2975(44\%)\cr
\hline
\hline
$W$+jet [fb] & 
16920& 4274(25\%)& 5791(34\%)& 8551(51\%)& 3063(18\%)& 
4521(27\%)& 7620(45\%)\cr
$\epsilon_{S/B}$          &0.77&  1.16	&	1.06	&	0.95	&	1.25	&	1.15	&	0.99\cr
$\epsilon_{S/\sqrt{B}}$&1.23& 0.94	&	1	&	1.08	&	0.86	&	0.95	&	1.06\cr
\hline
QCD [pb] & 
4402& 644(15\%)& 936(21\%)& 1627(37\%)& 337(8\%)& 
584(13\%)& 1279(29\%)\cr
$\epsilon_{S/B}$          &0.55&1.44	&	1.23	&	0.93	&	2.13	&	1.65	&	1.1\cr
$\epsilon_{S/\sqrt{B}}$&1.04& 1.04	&	1.07	&	1.07	&	1.11	&	1.14	&	1.12\cr
\hline
\end{tabular}
\caption{Numbers of tagged tops with $R=1.8$ with several cuts on
  $\Delta m^\text{prune}$ or $\Delta m^\text{unfilter}$.  The
  percentages are relative to the numbers of tagged tops without
  pruned or unfiltered mass cut in each category.  $\epsilon_{S/B}$
  and $\epsilon_{S/\sqrt{B}}$ denote improvement factors relative to
  the $R=1.5$ numbers with no cuts as shown in Tab.~\ref{tab:prune}.
  }
\label{tab:18prune}
\end{table}

The tagging efficiency for $R=1.8$ as a function of $p_{T,t}$ as well
as the fraction of hadronic tops included in a fat jet we show in the
left panel of Fig.~\ref{fig:18}.  By increasing $R$ from $1.5$ to $1.8$ we
increase the tagging efficiency for tops with $p_{T,t}=150-250$~GeV by
a factor 1.5 to 3. This is mainly an effect of more hadronic tops
fully included in the fat jet.  In the $300-450$~GeV range the effect
of an increased $R$ is small, and for $p_{T,t}>450$ the efficiency
even slightly decreases due to combinatorics.

The central panel shows the fractions of type~1 candidates and type~1
tags relative to all hadronic tops included in a fat jet as functions
of hadronic top $p_T$.  All the way down to $p_{T,t} = 100$~GeV the
efficiencies are flat, which means we have a fair chance to collect
very moderately boosted tops.  Type~2 and type~3 fractions we show at
the bottom of the figure.  While the fraction of type~2 and type~3
tags increases the fraction of type~1 candidates and tags does not
drastically change compared to $R=1.5$ as shown in the right panel
of Fig.~\ref{fig:had}.

Finally, increasing $R$ also increases the mis-tag rate significantly.
The right panel of Fig.~\ref{fig:18} shows the mis-tag efficiency as
functions of the fat jet $p_T$ for QCD and $W+$jets events. We observe
a larger increase for these background processes than for the signal,
which means we will not improve $S/B$ through larger jet
sizes. However, we might improve the statistical significance measure
$S/\sqrt{B}$.\bigskip

In Tab.~\ref{tab:18prune} we show the improvements in $S/B$ and
$S/\sqrt{B}$ relative to $R=1.5$ case.  We see that roughly twice the
number of tops get tagged, mainly at low transverse momenta. However,
$S/B$ still decreases by a factor $1/2$, while $S/ \sqrt{B}$ slightly
improves as long as QCD is the main background.

To compensate for the increased backgrounds we can also apply pruning
for $R=1.8$.  The corresponding efficiencies for different pruned mass
cuts we also include in Tab.~\ref{tab:18prune}. Adding pruning shifts
back the performance to a similar level as our $R=1.5$ results, and
there is no obvious advantage in combining it with larger fat jets,
unless there should be a specific reason to target the low-$p_{T,t}$
regime.

\section{Bottom tagging}
\label{sec:btag}

A major difference in the background rejection between the C/A based
Higgs tagger~\cite{bdrs} and corresponding top taggers are additional
$b$-tags.  Only based on kinematic conditions it appears unlikely to
achieve a QCD or $W$+jets rejection of more than a factor 1/100.
However, we can gain a significant improvement by requiring a $b$-tag
for one of the the top decay jets. At the same time, for moderately
boosted top quarks the kinematic tagging algorithm might benefit from
the identification of the $b$-jet, so we can first ensure that it is
captured and second use this information in the kinematic
reconstruction.\bigskip

Our first attempt to improve top tagging through an additional $b$-tag
will leave the kinematic top tagging algorithm unchanged and will
instead focus on the selection of the subjet which should be
$b$-tagged.  All of the usual top taggers treat three subjets
democratically, \ie without any $b$-tagging information.  We label the
$b$, $W_1$ and $W_2$ subjets such that $W_1$ and $W_2$-subjets
(ordered by hardness) reconstruct $m_W$ best; $j_b$, $j_{W1}$, and
$j_{W2}$ are the corresponding parton labels from Monte Carlo truth,
as defined in Sec.~\ref{sec:reco}. An obvious question is for what
fraction of all tags the subjet labeled $b$ really points to the
bottom quark $b=j_b$. For type~1 tags by definition one of the subjets
corresponds to a bottom, so $b=j_b$ implies that all subjets are
correctly assigned. For such tags a $W$ decay angle
analysis~\cite{hopkins} will work well.

\begin{table}[b]
\begin{tabular}{l||rr|rrrr}
\hline
&\multicolumn{2}{c|}{ all $n_W$} &  \multicolumn{2}{c}{$n_W=1$} & \multicolumn{2}{c}{$n_W=2$} \cr
\hline
$t\bar{t}$ [fb] & 14156&&8058&&6099\cr
$b=j_b$  & 9325&(66\%) & 5882&(73\%) &3442&(56\%)\cr
$b=j_{W1}$  & 2971&(21\%) &1242&(15\%) &1728&(28\%)\cr
$b=j_{W2}$  & 1666&(12\%) & 833&(10\%) &833&(14\%)\cr
\hline
type~1 & 10318&& 5808&& 4509\cr
$b=j_b$  &7917&(77\%) & 5044&(87\%) &2874&(64\%) \cr
$b=j_{W1}$  & 1695&(16\%)&502&(9\%) &1193&(26\%)\cr
$b=j_{W2}$  & 706&(7\%) &263&(4\%) &443&(10\%)\cr
\hline
type~2 & 1336&&781&&555\cr
$b=j_b$  &565&(42\%) & 341&(44\%) &224&(40\%) \cr
$b=j_{W1}$  & 499&(37\%) &294&(38\%) &205&(37\%)\cr
$b=j_{W2}$  & 392&(29\%) &226&(29\%) &166&(30\%)\cr
\hline
type~3 & 2503&&1468&&1035\cr
$b=j_b$  &842&(34\%) & 498&(34\%) &344&(33\%)\cr
$b=j_{W1}$  & 777&(31\%) & 447&(30\%) &331&(32\%)\cr
$b=j_{W2}$  & 568&(23\%) & 344&(23\%) &224&(22\%)\cr
\hline
$W$+jet [fb] &6590 & &3733& &2857\cr
QCD [pb] &1229 & &713& &516\cr
\hline
\end{tabular}
\caption{Kinematic identification probabilities ($b$) for all three
  top decay partons for different types of top tags and for different
  numbers of subjet pairings consistent with $m_W$ ($n_W$).
  }
\label{tab:btag}
\end{table}

In the first column of Tab.~\ref{tab:btag} we summarize the
probabilities to correctly assign the label $b$ in the
kinematics-based top tagging.  The percentages are defined relative to
the number of tagged tops in each category.  For type~1 tags there
will be exactly one $b$-parton in the tagged top while for for type~2
and type~3 tags there can be any number of $b$-partons. This means
that only for type~1 tags all three fractions sum to 100\%.  Almost
80\% of type~1 tags correctly assigns $b=j_b$, so they make a good
test sample for the question if identifying $b=j_b$ through a $b$-tag
leads to an efficient rejection for $W+$jets and QCD events.
From Tab.~\ref{tab:btag} we also see that the second-most likely
subjet to be kinematically identified as a $b$ really is
$j_{W1}$. However, this probability is less than a third of that for
$b=j_b$.  To rely on an additional $b$-tag for the case $b=j_{W1}$
does not even improve $S/\sqrt{B}$, so the only way to increase $S/B$
is to apply a $b$-tag only to the kinematically identified $j_b$.

As we show in Fig.~\ref{fig:2d_all}, a significant fraction of events
passing the mass plane cut have two pairs of subjets consistent
with the $W$ mass. This happens because the upper bound $m_{bj}^2 <
m_t^2 - m_W^2$ is numerically close to $m_W^2$~\cite{wjets}.
We expect that tagged tops where only one pair of subjets is
consistent with $m_W$ should more reliably give $b=j_b$. To test this,
we count the number of subjet pairs consistent with $m_W$ as $n_W=2$
when $\Delta m_2 - \Delta m_1 < 10\% \times m_W$ where $\Delta m =
|m_{jj} - m_W|$ for each pairing. Because $n_W=3$ essentially never
appears we find $n_W=1$ for the other top tags.

The distributions for kinematically identified $b$-subjets for
different $n_W$ we also show in Tab.~\ref{tab:btag}.  The fractions of
tagged tops with $n_W=1$ is almost the same ($\sim 57\%$) for
$t\bar{t}$ and QCD, $W+$jets.  Type~1 tagged tops with $n_W=1$ are
more likely to give $b=j_b$ than those with $n_W=2$, while for type~2
and type~3 we do not observe a significant difference.\bigskip

\begin{table}[b]
\begin{tabular}{c||r|rr|rr||rr}
\hline
& tagged & \multicolumn{2}{c|}{$b=j_b$} & \multicolumn{2}{c||}{$b=j_b$ and $n_W=1$} & 
\multicolumn{2}{c}{$b$-tag with cut for $m_{12}$}\cr
\hline
$t\bar{t}$ &1& 0.33 & (0.66$\varepsilon_b$)& 0.21 & (0.42$\varepsilon_b$)  & 0.41 & (0.82$
\varepsilon_b$) \cr
type~1 &0.73 & 0.28 & (0.56$\varepsilon_b$)   & 0.18 & (0.36$\varepsilon_b$)& 0.36 & (0.72$
\varepsilon_b$) \cr
\hline
$W$+jet &1& 0.02  & ($\varepsilon_b^\text{mis}$) & 0.0114  & (0.57$\varepsilon_b^\text{mis}$)& 
0.0318 & (1.59$\varepsilon_b^\text{mis}$)  \cr
QCD &1 & 0.02 & ($\varepsilon_b^\text{mis}$) & 0.0116 & (0.58$\varepsilon_b^\text{mis}$) & 
0.0332  & (1.66$\varepsilon_b^\text{mis}$  )\cr
\hline\hline
$\epsilon_{S/B}$ &1& 16.5   & (0.66$\varepsilon_b/\varepsilon_b^\text{mis}$)& 18.42 & 
(0.72$\varepsilon_b/\varepsilon_b^\text{mis}$) & 12.35 & (0.49$\varepsilon_b/\varepsilon_b^\text{mis}$) 
\cr
$\epsilon_{{S/\sqrt{B}}}$ &1 & 2.33 & (0.66$\varepsilon_b/{\sqrt{\varepsilon_b^\text{mis}}}$)  
&1.95& (0.55$\varepsilon_b/\sqrt{\varepsilon_b^\text{mis}}$)& 2.25 & (0.64$\varepsilon_b/
\sqrt{\varepsilon_b^\text{mis}}$) \cr
\hline
\end{tabular}
\caption{Efficiencies of a $b$-tag after top tagging and for the
  modified tagger with $b$-tagging.  We assume $\varepsilon_b=0.5$ and
  $\varepsilon_b^\text{mis}=0.02$ and quote the improvement factors
  $\epsilon_{S/B}, \epsilon_{S/\sqrt{B}}$ against the QCD background.}
\label{tab:btag2}
\end{table}

Consequently, the best strategy to improve $S/B$ in top tagging is to
target $n_W=1$ events and check the kinematically identified
$b$-subjet with a $b$-tag. In terms of a tagging efficiency
$\varepsilon_b$ and a mis-identification rate
$\varepsilon_b^\text{mis}$ for light flavors (we ignore $c$ quarks in
this simple estimate) we show the possible improvements in
Tab.~\ref{tab:btag2}.  We expect an enhancement by
$0.66\varepsilon_b/\varepsilon_b^\text{mis}$ ($0.73\varepsilon_b/
\varepsilon_b^\text{mis}$ for selecting $n_W=1$) for $S/B$, so
assuming $\varepsilon_b=50\%$ and $\varepsilon_b^\text{mis}=2\%$ we
find that $S/B$ improves by 16.5 (18 for selecting
$n_W=1$). Similarly, we find an improvement in $S/\sqrt{B}$ around a
factor 2.\bigskip

\begin{figure}[t]
\includegraphics[width=0.3\textwidth]{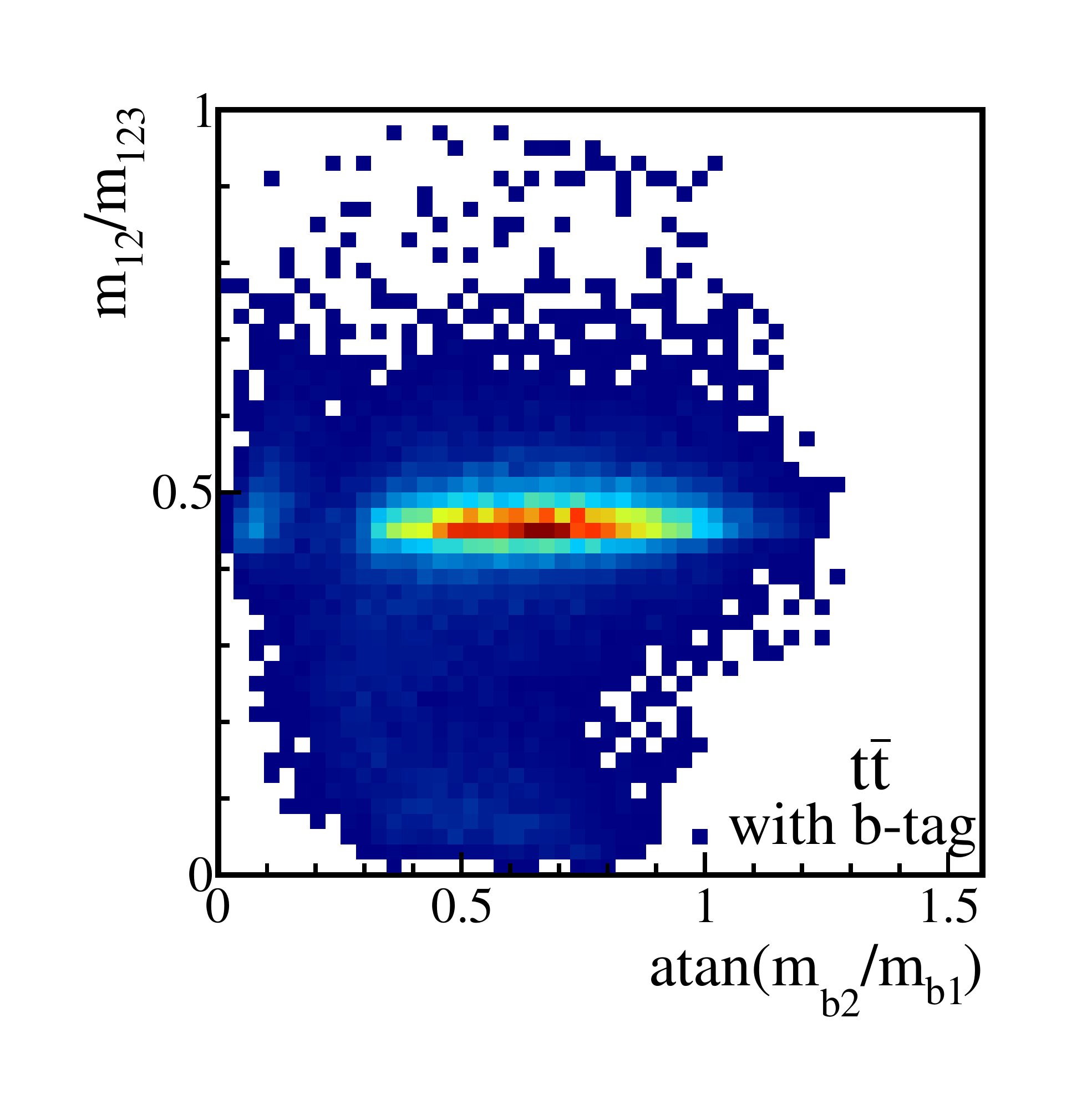}
\includegraphics[width=0.3\textwidth]{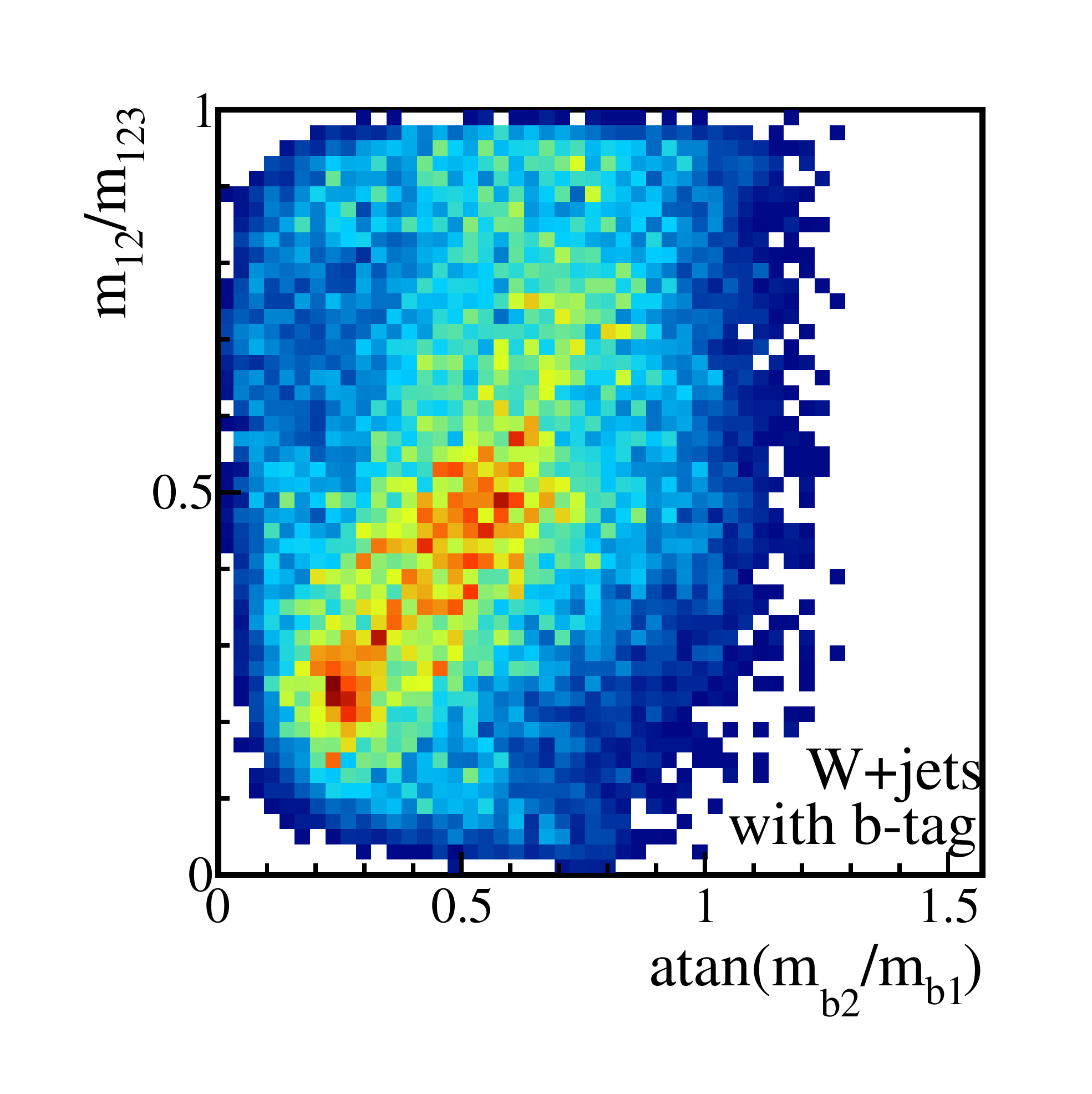}
\includegraphics[width=0.3\textwidth]{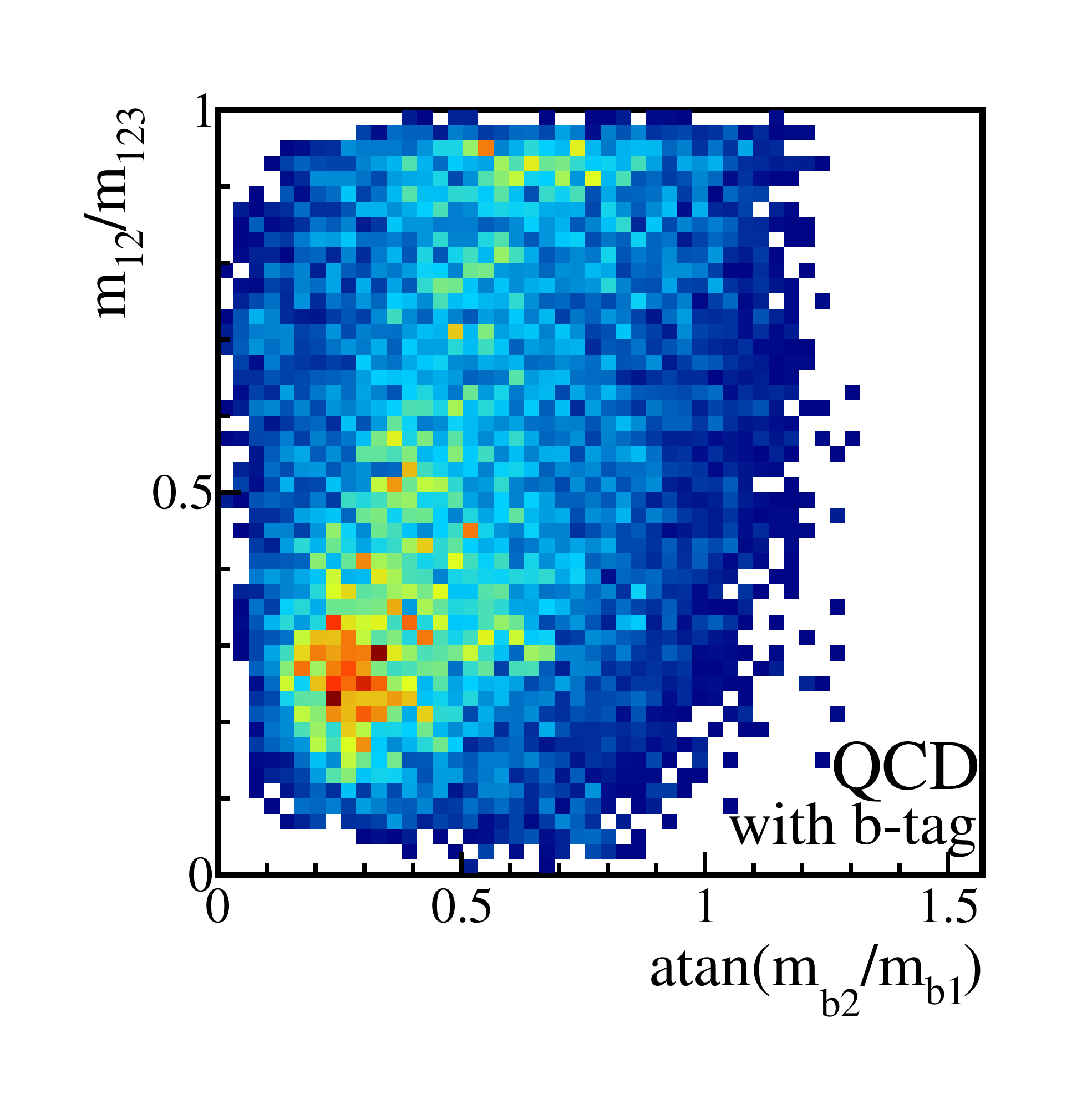}
\vspace*{-4mm}
\caption{ Modified mass plane distribution for signal, $W+$jets and
  QCD events.  The mass plane is now defined as
  $\arctan(m_{b2}/m_{b1})$ vs. $m_{12}/m_{123}$.  }
\label{fig:btag2d}
\end{figure}

As an alternative to an external $b$-tag, we can use the $b$-tagging
information during the kinematic top tagging algorithm. The
selection of the three decay subjets after the mass drop criteria we
modified in four steps:
\begin{enumerate}
\item group the subjets into either $b$-tagged or 
  non-$b$-tagged subjets.
\item take all possible triplets of one $b$-tagged and two
  non-$b$-tagged subjets and select the one with the best filtered top
  mass.
\item check if this $m^\text{filter}$ satisfies our criterion.
\item apply modified mass plane cuts.
\end{enumerate}
Since we require exactly one $b$-tagged subjet in step~2, we adapt the
labels $b$, $W_1$ $W_2$ to reflect this.  In step~4, we should change
the invariant masses in terms of $(p_1, p_2, p_3)$ into $(p_b, p_{W1},
p_{W2})$. Our two-dimensional mass plane becomes
$\arctan(m_{b2}/m_{b1})$ vs. $m_{12}/m_{123}$, where $m_{bi}^2=(p_b +
p_{Wi})^2$ and $m_{12}^2=(p_{W1} + p_{W2})^2$.  Fig.~\ref{fig:btag2d}
shows the two-dimensional distribution of top candidates in the
modified mass plane for semi-leptonic $t\bar{t}$ events, $W+$jets, and
QCD events. Unlike before, the signal events now show a clear $W$ mass
peak only for $m_{12}$. To see how well this new algorithm might do we
assume a 100\% $b$-tagging efficiency of perfect purity. Because for
the backgrounds all three subjets could equally likely be mis-tagged
we simply reweighting each of the possibilities by
$\varepsilon_b^\text{mis}$.  Following these plots we can apply
stricter mass plane cut on the modified mass plane than before.  For
our test we use
\begin{alignat}{5}
\left|\frac{m_{12}}{m_{123}} -\frac{m_W}{m_t}\right| < 15\% 
\qqquad \text{and} 
\qquad 
0.2< \arctan \left(\frac{m_{b2}}{m_{b1}}\right)<1.3 \; .
\label{eq:cut}
\end{alignat}
\bigskip

\begin{figure}[b]
\includegraphics[width=0.3\textwidth]{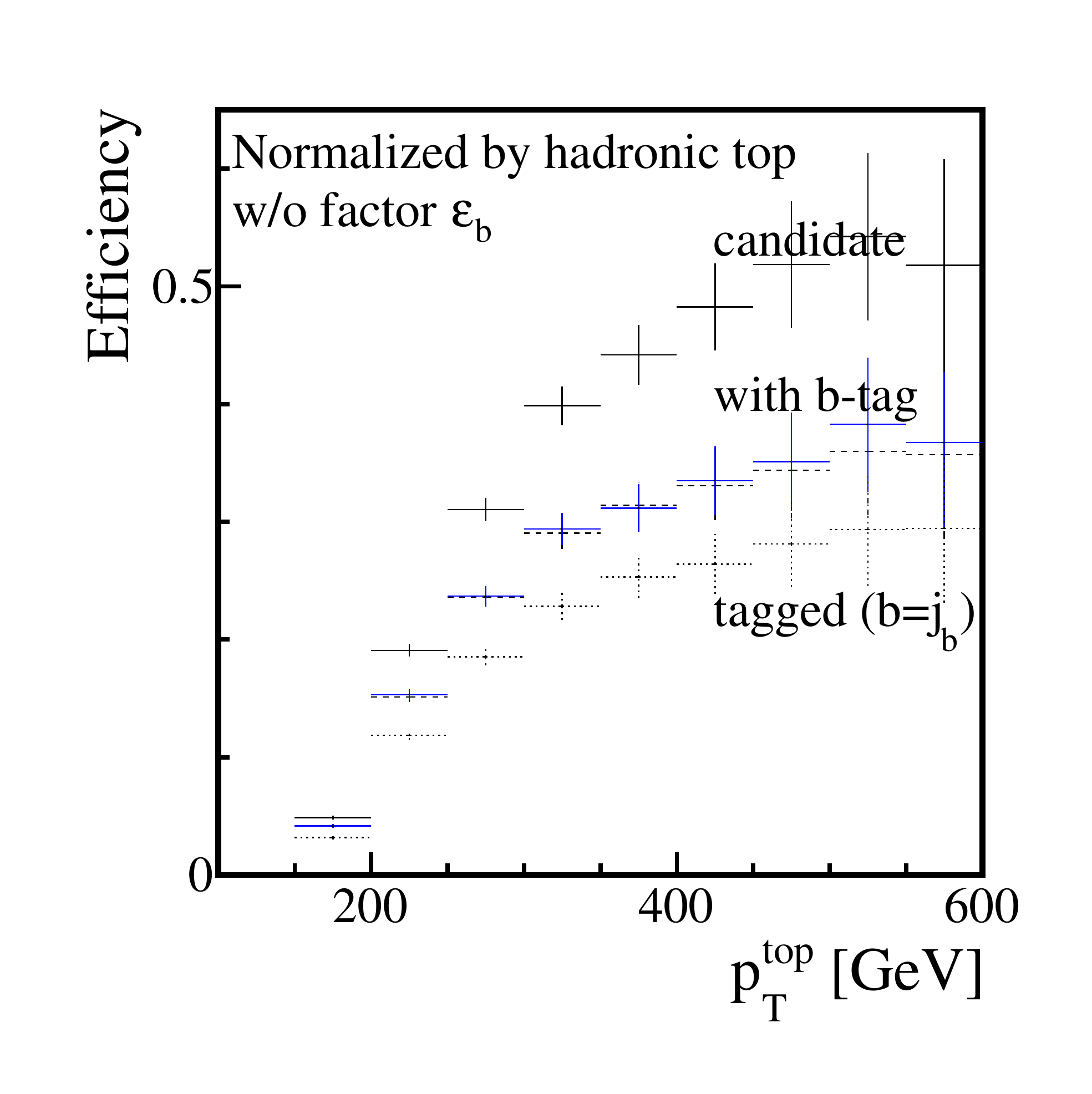}
\includegraphics[width=0.3\textwidth]{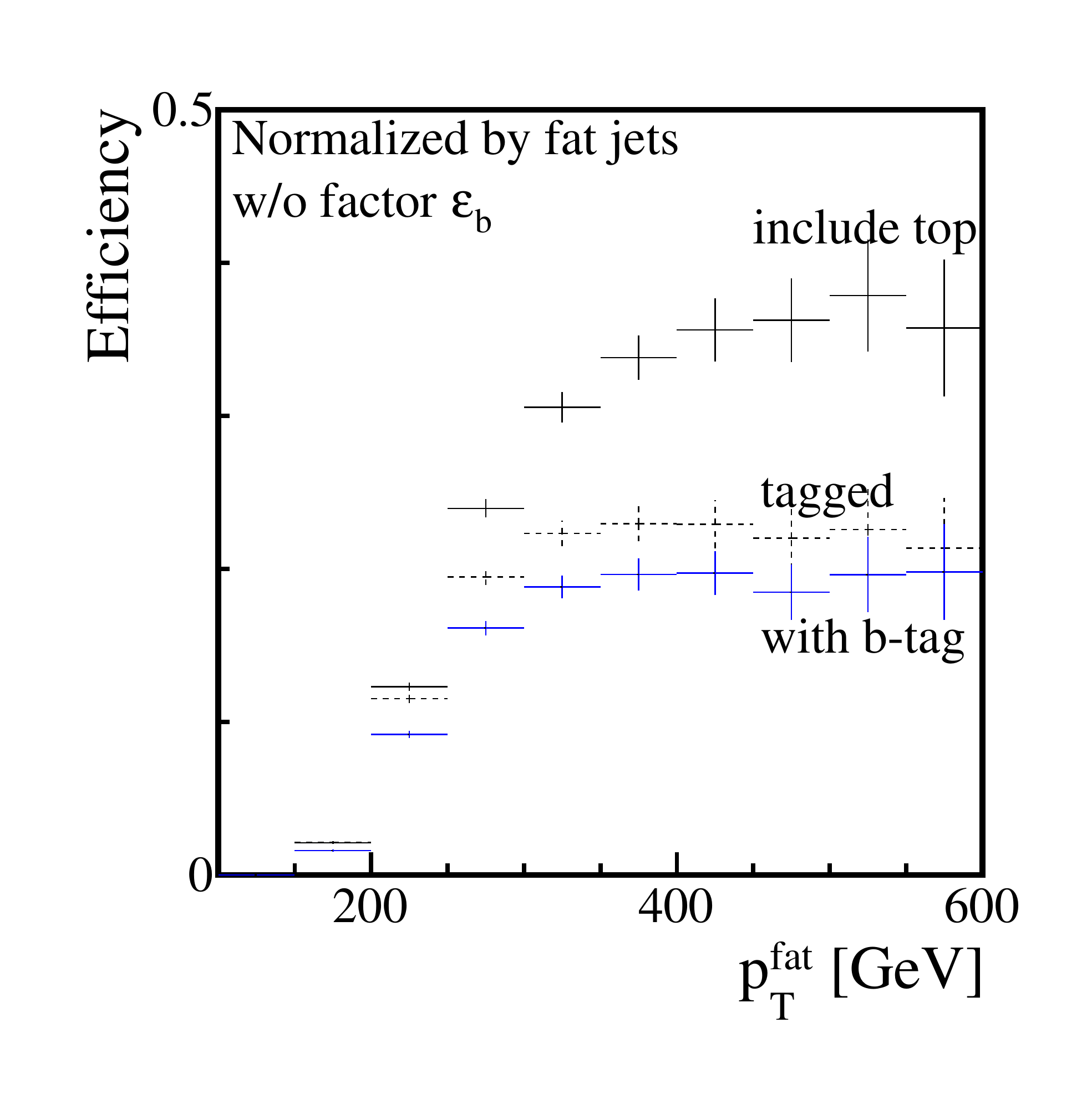}
\includegraphics[width=0.3\textwidth]{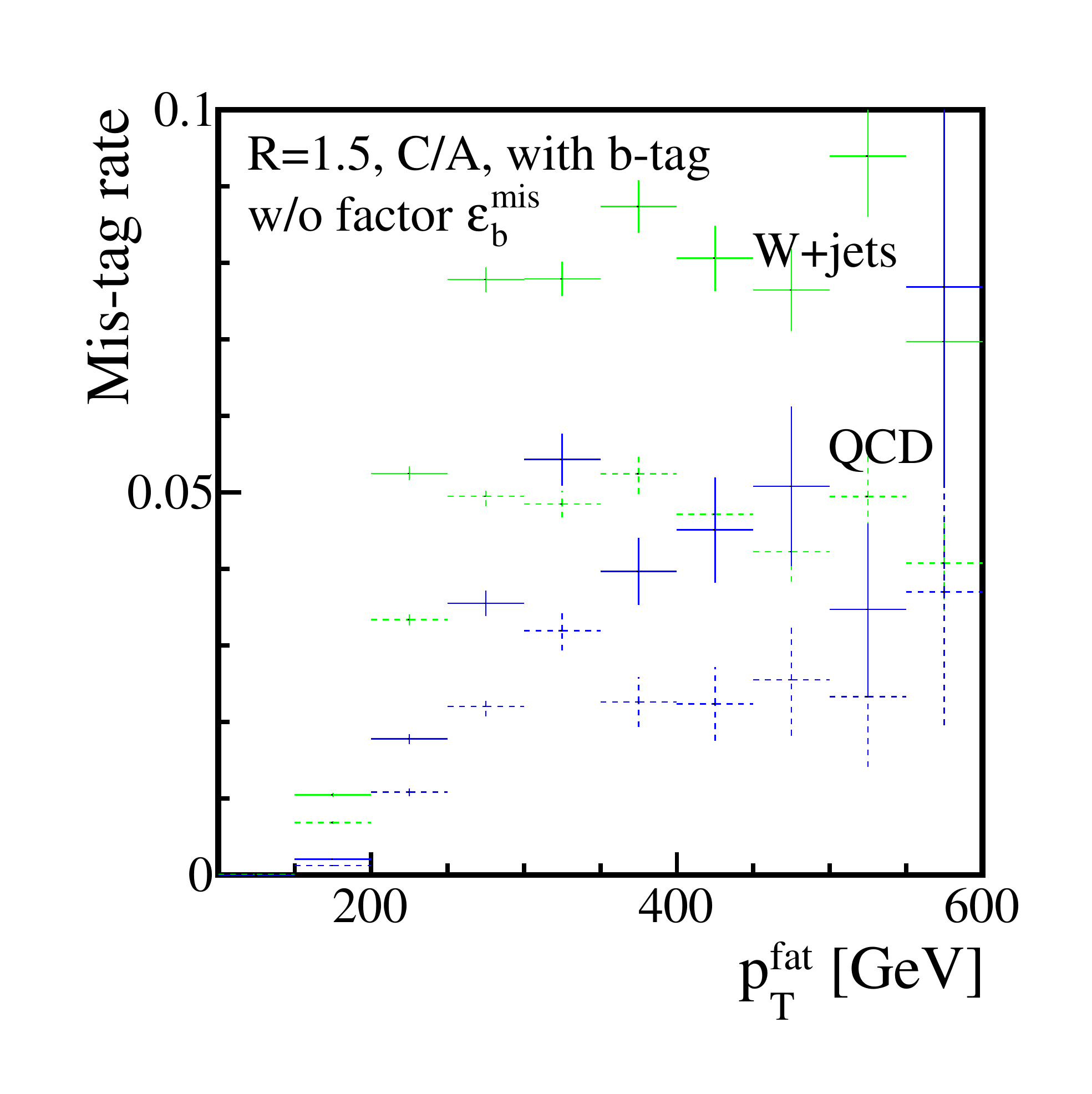}
\vspace*{-4mm}
\caption{ Left: tagging efficiency of the modified algorithm as a
  function of $p_{T,t}$ and assuming $\varepsilon_b=1$.  Center:
  tagging efficiency of the modified algorithm as a function of the
  $p_T$ of the fat jet.  Right: mis-tagging rate of the modified
  algorithm as a function of the $p_T$ of the fat jet for QCD (blue)
  and $W+$jets samples (green).  Again, we omit
  $\varepsilon_b^\text{mis}$.  In all panels dotted lines show the
  {\sc HEPTopTagger} results as a reference.}
  \label{fig:btageff}
\end{figure}

Based on this modified algorithm the left panel of
Fig.~\ref{fig:btageff} shows the signal efficiency as a function of
$p_{T,t}$. Again, for the signal we assume a perfect $b$-tag.  We can
simply multiply by $\varepsilon_b$ to compute the final top tagging
efficiency, ignoring the small effect of $\varepsilon^\text{mis}_b$.
Similarly, we ignore cases with two $b$-subjets in the tagged top as
subleading by a factor $\varepsilon_b (1-\varepsilon_b)$ and only
appearing for type~2 or type~3.  

Taking into account the probability of 77\% with which the default top
tagger correctly assigns $b=j_b$, the modified algorithm slightly
increases the number of tagged tops if we assume perfect
$b$-tagging. It returns almost the same tagging efficiency as the
purely kinematic tagger before adding the factor $\varepsilon_b$.
This is because the default top tagger identifies the crucial type~1
tops even without $b$-tagging information while type~2 or type~3
configurations are comparably rare. The use of $b$-tagging really only
helps to identify which subjet of a type~1 tag corresponds to the
bottom and to reject type~2 and type~3 tags.  To confirm this, in the
central panel of Fig.~\ref{fig:btageff} we show the efficiency as a
function of the fat jet's $p_T$. Compared to $p_{T,t}$ this
effectively removes some type~2 and type~3 tops.

The right panel of Fig.~\ref{fig:btageff} shows the mis-tag rate of
the modified algorithm.  Here we simply assume that one of the three
subjets found by the kinematic top tagger is mis-identified.  These
candidate tops have to be multiplied by $3 \times \varepsilon_b^\text{mis}$
before imposing the modified mass plane cut.  The effect of
mis-identifying $b$-subjets among others than the three subjets
selected by the kinematic tagger we neglect. It should be less than
10\% given that more than 90\% of the fat jets in which three subjets
fulfill the top mass criteria only have one such combination.

The main question is if the restricted mass plane cuts now reduce the
backgrounds more efficiently.  We find that the cuts from
Eq.(\ref{eq:cut}) efficiently drop tagged tops from background samples
as compared to democratic mass plane cuts, but they do not compensate for the 
combinatorial factor 3 in the mis-tagging probability.\bigskip

\begin{figure}[t]
\includegraphics[width=0.3\textwidth]{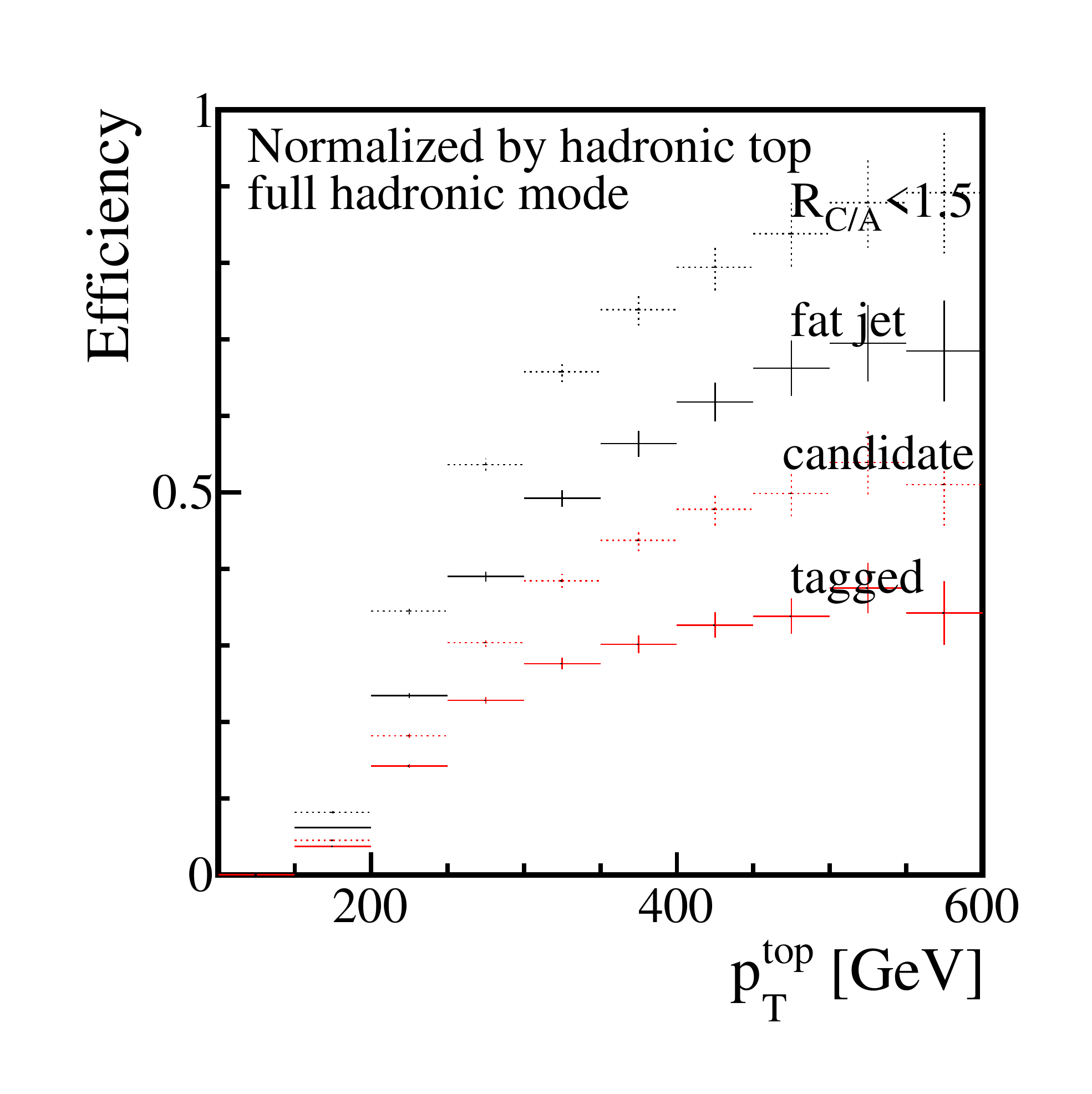}
\includegraphics[width=0.3\textwidth]{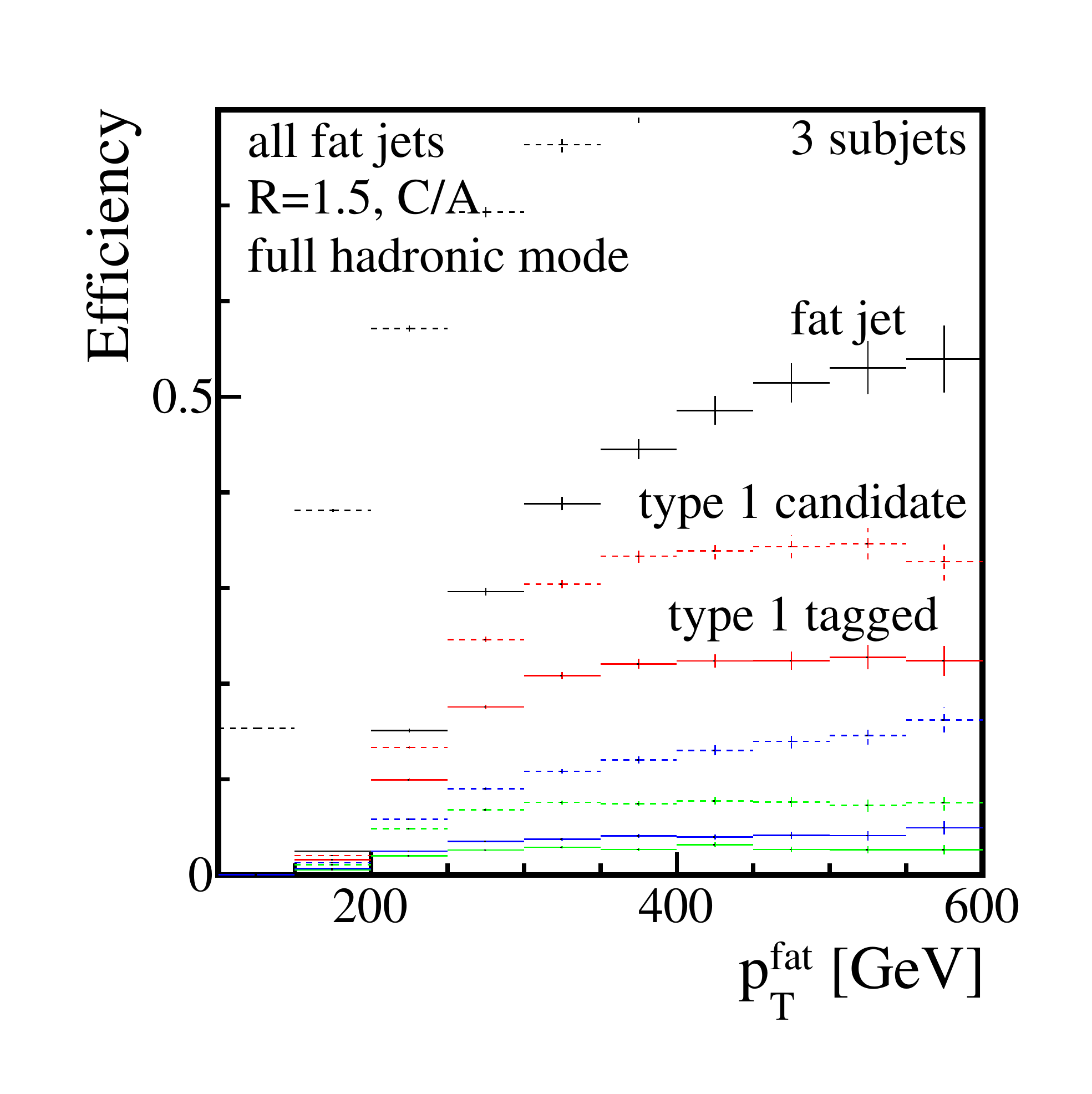}
\includegraphics[width=0.3\textwidth]{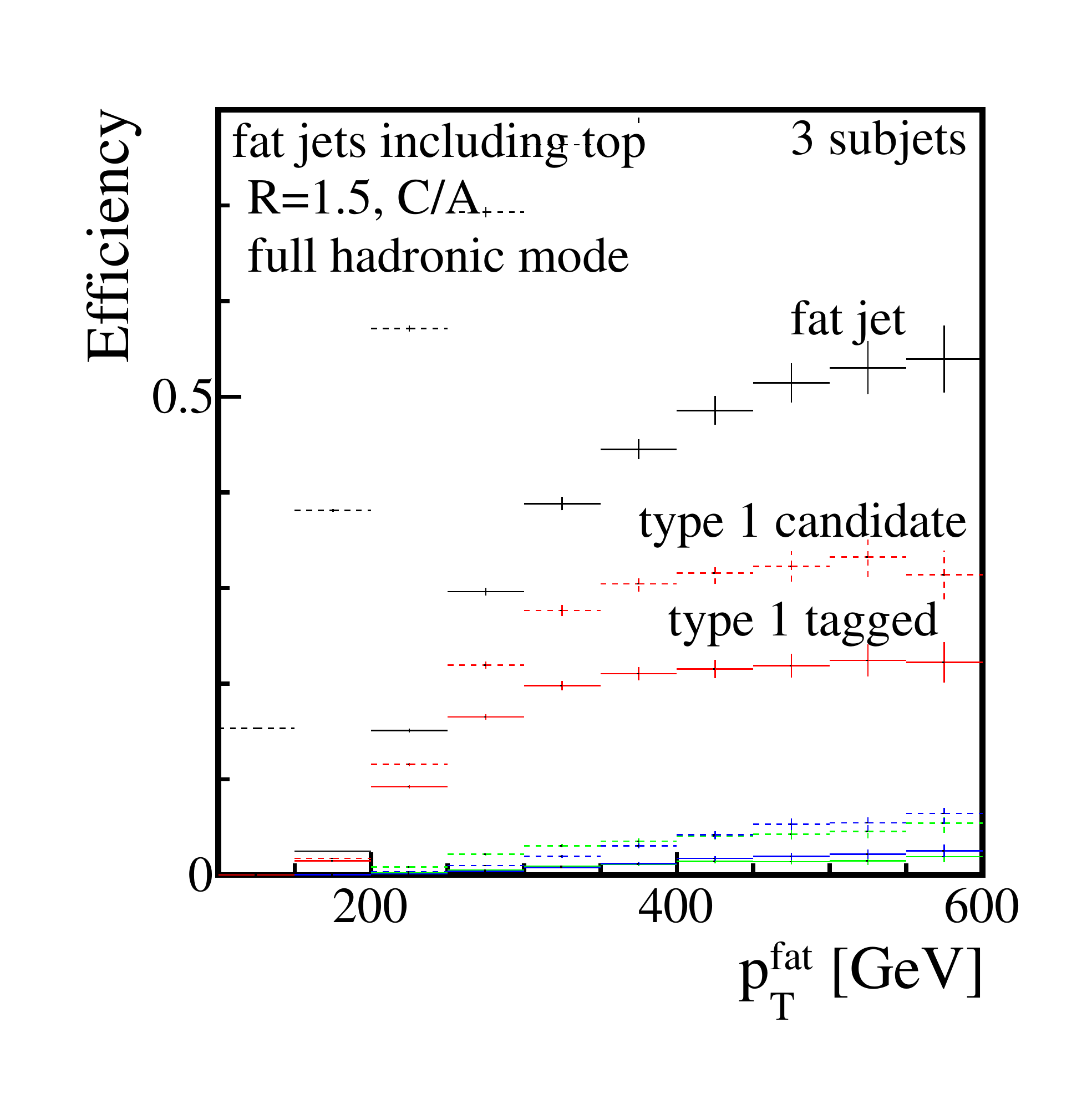}
\vspace*{-4mm}
\caption{Left: tagging efficiencies $\epsilon_{t\bar{t}}$ normalized
  to the number of hadronic tops as a function of the $p_T$ for the
  fully hadronic $t\bar{t}$ sample.  Center: efficiencies for type~1
  (red), type~2 (green) and type~3 (blue) as functions of the fat jet
  $p_T$ for the fully hadronic $t\bar{t}$ sample.  The dotted lines
  show the corresponding candidate fractions. Right: fraction of
  type~1, type~2 and type~3 only for fat jets including a top.}
\label{fig:full}
\end{figure}

Up to this point we have only considered semi-leptonic $t\bar{t}$
events.  In particular for the possible improvement through
$b$-tagging we should consider the more combinatorics prone purely
hadronic decays.  The left panel of Fig.~\ref{fig:full} shows the
tagging efficiencies as a function of $p_{T,t}$ for the fully hadronic
$t\bar{t}$ sample. Compared to Fig.~\ref{fig:had} the tagger works with
almost the same efficiency, provided we normalize the number of tagged
events to all hadronic tops.

The central panel of Fig.~\ref{fig:full} shows the default top tagger
efficiencies as a function of the transverse momentum of the fat jet,
again for hadronic decays. The tagging efficiency and the fraction of
candidate tops we show for type~1, type~2 and type~3 tags. Since the
number of fat jets including a hadronic top shown as the black solid
line increases compared to semi-leptonic events the resulting
candidate and tagged efficiency becomes larger. The number of type~2
and type~3 tags increases simply with the jet multiplicities. In the
right panel of Fig.~\ref{fig:full} we show the same fractions as in
the central panel but requiring that all fat jets include a hadronic
top. For such fat jets we find type~2 or type~3 almost as rarely as in
the semi-leptonic sample, \ie type~2 and type~3 candidates contribute
at most 30\% relative to type~1 candidates.  This means that even for
the fully hadronic $t\bar{t}$ sample a modified algorithm including
$b$-tagging will not provide enough of an improvement to compensate an
expected factor 50\% increase in the mis-tag rate.\bigskip

In summary, using $b$-tagging when selecting the relevant three
subjets in a fat jet does not enhance $S/B$. The purely kinematic {\sc
  HEPTopTagger} selects the correct set of subjets too reliably to
gain a significant improvement as long as the hadronic top is fully
captured in a fat jet. The $b$-subjet selected based on kinematics is
usually identified correctly.  Relying on $b$-tagging inside the
kinematic algorithm is hurt by the combinatorial mis-tagging
efficiency of $3 \times \epsilon^\text{mis}$ while there is no such
factor 3 for the signal. This disadvantage is hard to compensate by
improved mass plane cuts.  To improve $S/B$, the best approach is to
use the $b$-tag only for the most probable $b$-subjet and simply add
it to the kinematic tagging algorithm.

\section{Outlook}
\label{sec:outlook}

In this paper we have proposed and tested several modifications to
kinematic top tagging, as implemented in the {\sc
  HEPTopTagger}. As a starting point we have shown that provided the
top quark is boosted enough to be collected inside a fat jet the usual
kinematic criteria, \ie a search for mass drops in the clustering
history and the reconstruction of three independent invariant mass
variables from the suspected top decay subjets, do not exhibit obvious
shortcomings.

\begin{enumerate}
\item One possible improvement, curing for example the decreasing
  efficiency of C/A based taggers towards larger boost is a switch to
  the $k_T$ algorithm once we identified the main subjets using mass
  drops. It keeps the tagging efficiency relative to the number of
  tops caught inside a fat jet on a plateau over the entire range
  $p_{T,t} \sim 150- 600$~GeV.

\item Using pruning in combination with the usual filtering procedure
  we gain an additional kinematic variable. Cutting on it can almost
  double $S/B$ relative to pure QCD backgrounds. However, this
  improvement should be taken with a grain of salt until it can be
  confirmed by a proper detector simulation in the presence or pile-up.

\item To extend the top tagger to lower boost we can increase the
  original size of the fat jet from $R=1.5$ to $R=1.8$. Indeed, the
  efficiency for low transverse momenta increases, but as well does the
  background combinatorics. We find that neither $S/B$ nor
  $S/\sqrt{B}$ strongly benefit from this modification.

\item Including a $b$-tag as an additional step in the top tagging
  procedure can very significantly enhance the background
  rejection. This is well known from Higgs tagging. We also find that
  for the {\sc HEPTopTagger} including such a $b$-tag in a modified
  algorithm is not promising.
\end{enumerate}

At a time when we can expect the first officially tagged top quarks at
the LHC these studies can guide us towards possible modifications and
improvements of top taggers with different analyses in mind. They show
that for example in the case of the {\sc HEPTopTagger} there is still
room for adjustment but that top taggers in general have within a few
years reached an impressive level of maturity and reliability.

\begin{acknowledgments}
We would like to thank our experimental ATLAS colleagues Dirk Zerwas,
Sebastian Sch\"atzel, and in particular Gregor Kasieczka for their
support and advice. Without personal contacts progress in this field
of LHC phenomenology would not be possible. Felix Kling we would like to
thank for carefully checking our analysis code.
\end{acknowledgments}


\end{document}